\documentclass[a4paper,fleqn,usenatbib]{mnras}

\usepackage{newtxtext,newtxmath}

\usepackage[T1]{fontenc}
\usepackage{ae,aecompl}

\usepackage{graphicx}	% Including figure files

\newcommand{\spose}[1]{\hbox  to 0pt{#1\hss}}  
\newcommand{\lta}{\mathrel{\spose{\lower 3pt\hbox{$\sim$}}\raise  2.0pt\hbox{$<$}}}
\newcommand{\gta}{\mathrel{\spose{\lower  3pt\hbox{$\sim$}}\raise 2.0pt\hbox{$>$}}}
\newcommand \HI {\ifmmode \rm {\sc Hi} \else {\sc Hi}\fi}
\newcommand{\kms} {\ifmmode  \,\rm km\,s^{-1} \else $\,\rm km\,s^{-1}  $ \fi }
\newcommand{\kpc} {\ifmmode  {\rm kpc}  \else ${\rm  kpc}$ \fi  }  
\newcommand{\Msun} {\ifmmode \rm M_{\odot} \else $\rm M_{\odot}$ \fi}

\newcommand{\LCDM} {\ifmmode \Lambda{\rm CDM} \else $\Lambda{\rm CDM}$ \fi}
\newcommand{\Omegam} {\ifmmode \Omega_{\rm m} \else $\Omega_{\rm m}$ \fi} 
\newcommand{\Omegab} {\ifmmode \Omega_{\rm b} \else $\Omega_{\rm b}$ \fi} 
\newcommand{\OmegaL} {\ifmmode \Omega_{\rm \Lambda} \else $\Omega_{\rm \Lambda}$\fi} 
\newcommand{\rhocrit} {\ifmmode \rho_{\rm crit} \else $\rho_{\rm crit}$ \fi}

\newcommand{\Vmaxdmo} {\ifmmode V_{\rm  max}^{\rm DMO} \else  $V_{\rm max}^{\rm DMO}$  \fi} 
\newcommand{\Mhalo} {\ifmmode M_{\rm halo} \else $M_{\rm  halo}$ \fi}  

\newcommand{\Vflat} {\ifmmode V_{\rm flat} \else $V_{\rm flat}$ \fi}
\newcommand{\VHI} {\ifmmode V_{\rm HI} \else $V_{\rm HI}$ \fi} 
\newcommand{\Mstar} {\ifmmode M_{\rm star} \else $M_{\rm star}$ \fi} 
\newcommand{\Mbar} {\ifmmode M_{\rm bar} \else $M_{\rm bar}$ \fi} 

\newcommand{\alphamin} {\ifmmode \alpha_{\rm min} \else $\alpha_{\rm min}$ \fi}

\title[Galactic uniformity] {NIHAO XII: galactic uniformity in a \LCDM universe}

\author[Dutton et al.]{Aaron  A. Dutton,$^{1}$\thanks{dutton@nyu.edu}
  Aura Obreja,$^1$ Liang Wang,$^{2,3}$ Thales A. Gutcke,$^4$ Tobias Buck,$^4$
  \newauthor{Silviu M. Udrescu,$^1$ Jonas Frings,$^4$ Gregory S. Stinson,$^4$ Xi Kang$^2$}
  \newauthor{and Andrea V. Macci\`o$^{1,4}$}
  \\
$^1$New York University Abu Dhabi, PO Box 129188, Abu Dhabi, United Arab Emirates\\
$^2$Purple Mountain Observatory, 2 West Beijing Road, Nanjing 210008, China\\
$^3$International Centre for Radio Astronomy Research (ICRAR), M468, University of Western Australia, 35 Stirling Hwy, Crawley,\\ WA 6009, Australia\\
$^4$Max-Planck-Institut f\"ur Astronomie, K\"onigstuhl 17, 69117 Heidelberg, Germany\\
}

\date{Accepted 2017 February 19. Received February 19; in original form 2016 October 20}

\pubyear{2017}

\begin{document}
\label{firstpage}
\pagerange{\pageref{firstpage}--\pageref{lastpage}}
\maketitle

\begin{abstract}
  We use a sample of 83 high-resolution cosmological zoom-in
  simulations and a semi-analytic model (SAM) to study the
  stochasticity of galaxy formation in haloes ranging from dwarf to
  Milky Way masses.  Our simulated galaxies reproduce the observed
  inefficiency of galaxy formation as expressed through the stellar,
  gas and baryonic Tully-Fisher relations.  For \HI\, velocities in the
  range ($70 \lta V \lta 220$ km/s), the scatter is just 0.08 to 0.14
  dex, consistent with the observed intrinsic scatter at these
  scales. At low velocities ($20 \lta V \lta 70$ km/s), the simulated
  scatter is 0.2-0.25 dex, which could be tested with future
  observations.  The scatter in the stellar mass versus dark halo
  velocity relation is constant for $30 \lta V \lta 180 \kms$, and
  smaller ($\simeq 0.17$ dex) when using the maximum circular velocity
  of the dark matter only simulation, \Vmaxdmo, compared to the virial
  velocity ($V_{200}$ or $V_{200}^{\rm DMO}$).  The scatter in stellar
  mass is correlated with halo concentration, and is minimized when
  using a circular velocity at a fixed fraction of the virial radius
  $\simeq 0.4 R_{200}$ or with $V_{\alpha}=V_{200}^{\rm DMO} (V_{\rm
    max}^{\rm DMO}/V_{200}^{\rm DMO})^\alpha$ with $\alpha\simeq 0.7$,
  consistent with constraints from halo clustering.  Using the SAM we
  show the correlation between halo formation time and concentration
  is essential in order to reproduce this  result.  This  uniformity
  in galaxy formation efficiency we see in our hydrodynamical
  simulations and a semi-analytic model proves the simplicity and
  self-regulating nature of galaxy formation in a $\Lambda$ Cold Dark
  Matter ($\Lambda$CDM) universe. 
\end{abstract}

\begin{keywords}
methods: numerical -- galaxies: fundamental parameters -- galaxies: haloes -- galaxies: kinematics and dynamics -- dark matter
\end{keywords}

\setcounter{footnote}{1}

%%%%%%%%%%%%%%%%%%%%%%%%%%%%%%%%%%%%%%%%%%%%%%%%%%%%%%%%%%%%%%%%%%%%%%
%% SECTION 1: INTRODUCTION
%%%%%%%%%%%%%%%%%%%%%%%%%%%%%%%%%%%%%%%%%%%%%%%%%%%%%%%%%%%%%%%%%%%%%%

\section{Introduction}
\label{sec:intro}

Galaxies come in a wide variety of sizes, shapes, and masses, yet
their structural properties obey a number of scaling relations --
hinting at an underlying simplicity to the seemingly haphazard process
of galaxy formation in a hierarchical universe.  The correlation
between dynamics and luminosity is fundamental, due to its small
scatter, and the link it provides between baryons and dark
matter. This relation is known as the Tully-Fisher
\citep[TF;][]{Tully77} and Faber-Jackson \citep[FJ;][]{Faber76}
relation, for spiral/late-type/star forming and
elliptical/early-type/quiescent galaxies, respectively. In the
original studies the dynamics was traced by the linewidth of the 21cm
line of neutral hydrogen (\HI) for spiral galaxies, and by the stellar
velocity dispersion for elliptical galaxies.  In more recent studies,
luminosity has been replaced by stellar or baryonic (defined as stars
plus neutral gas) mass \citep[e.g.,][]{McGaugh00,Bell01}, with these
relations often being referred to as the stellar and baryonic
Tully-Fisher (BTF) relations, respectively. In order to put all types of
galaxies on to the same relation circular velocities measured at the
same fiducial radius can be  used \citep{Dutton10b, Dutton11}.

The TF relation is a benchmark for any successful theory of galaxy
formation. The slope naturally arises in cold dark matter (CDM) based
galaxy formation models \citep[e.g.,][]{Mo98,Navarro00,Dutton07}, but
reproducing the normalization has been more challenging, with models
invariably predicting higher rotation velocities at fixed stellar mass
than observed \citep[e.g.][]{Navarro00, Governato07, Marinacci14}. A
contributor to this problem is when simulated galaxies are too
compact.  However, \citet{Dutton07,Dutton11} showed that when models
are constrained to reproduce the sizes of galaxies, they also
overpredict the circular velocities if haloes contract in response to
galaxy formation \citep{Blumenthal86, Gnedin04}. Such a conclusion has
been verified by subsequent studies \citep[e.g.,][]{Desmond15,Chan15}.

A close relative of the TF relation is that between galaxy
stellar mass and dark halo mass. The \Mstar versus \Mhalo relation has
been studied extensively using observations of weak gravitational
lensing, satellite kinematics and halo abundance matching
\citep[e.g.,][]{Yang03, Mandelbaum06, Conroy09, Moster10, More11,
  Leauthaud12, Behroozi13, Hudson15}. One of the key
results of these studies is that star formation is inefficient. The
maximum efficiency is $\approx 25\%$, and independent of redshift,
occurring at a halo mass similar to that of the Milky Way (i.e., $\sim
10^{12}\Msun$). At both higher and lower halo masses, the efficiency
drops even further.

The scatter in stellar mass at fixed halo mass is hard to quantify
observationally (since halo masses cannot be reliably measured for
individual galaxies), but it is estimated to be small, $\approx 0.20$ dex,
and independent of halo mass \citep{More11, Reddick13}. Note that the
scatter in halo mass at fixed stellar mass is not a constant, rather
it increases with stellar mass due to the shallowing slope of the
$\Mstar-\Mhalo$ relation at high halo masses.  Likewise, the observed
scatter in the TF relation is small $\approx 0.20$ dex in mass (or
luminosity) at fixed velocity \citep{Courteau07, Reyes11, McGaugh12}.
The maximum scatter in the stellar-to-halo mass ratio allowed by the
TF relation is slightly larger than the observed scatter, since
variation in stellar-to-halo mass partially moves galaxies along the
TF relation \citep{Dutton07}.  However, contributions from variation
in dark halo concentrations, stellar mass-to-light ratios (both
intrinsic variations and measurement uncertainties) reduce the allowed
scatter to less than 0.2 dex.

The best tracer of halo masses for individual galaxies comes from
galaxy rotation velocities at large galactic radii from the 21 cm line
of neutral hydrogen.  The BTF relation is a correlation between the
baryonic mass of a galaxy, $\Mbar$, (stars plus neutral gas) and the
rotation velocity at large radii, $\Vflat$, (typically in the flat
part of the rotation curve). It is an extension of the original
linewidth - luminosity relation of \citet{Tully77}. Since the observed
rotation velocity is related to the halo virial velocity
\citep{Dutton10b}, the BTF provides the most direct observational
constraint on the efficiency of galaxy formation.  The observed
scatter in the BTF is $\simeq 0.22$ dex in baryonic mass at fixed
rotation velocity \citep{McGaugh12, Lelli16}. Accounting for
measurement uncertainties the intrinsic scatter is $\simeq 0.10$ dex.
Using a semi-analytic model \citep[SAM;][]{Dutton09} for disc galaxy
formation, \citet{Dutton12} found a model scatter of $\simeq 0.15$
dex. Since this model makes some simplifying assumptions (such as
smooth mass accretion histories), one might expect it to underestimate
the true scatter.  Thus, there is a potential conflict between the
observed and predicted scatter in the BTF. 

Galaxy formation is observed to be both inefficient, yet remarkably
uniform.  Is it possible to reproduce this in a $\Lambda$ cold dark
matter ($\Lambda$CDM) structure formation scenario?  Which definition
of galaxy mass and halo mass are most tightly correlated?  To address
these questions, we study the relations between galaxy mass and
circular velocity in a new sample of high-resolution cosmological
simulations ($\sim 10^6$ particles per galaxy) from the Numerical
Investigation of a Hundred Astrophysical Objects (NIHAO) project
\citep{Wang15}. This paper is organized as follows: The NIHAO
simulations and derived galaxy properties are described in
\S\ref{sec:sims}, the velocity mass relations are shown in
\S\ref{sec:results}, implications for halo abundance matching are
shown in \S\ref{sec:ham}, and a summary is given in \S\ref{sec:sum}.

%%%%%%%%%%%%%%%%%%%%%%%%%%%%%%%%%%%%%%%%%%%%%%%%%%%
\section{Galaxy Formation Simulations} \label{sec:sims}
%%%%%%%%%%%%%%%%%%%%%%%%%%%%%%%%%%%%%%%%%%%%%%%%%%%

Here we briefly describe the NIHAO simulations. We refer the reader to
\citet{Wang15} for a complete discussion.  NIHAO  is a sample of
$\sim$ 90 hydrodynamical cosmological zoom-in simulations using the
smoothed particle hydrodynamics code {\sc gasoline} \citep{Wadsley04}
with improvements to the hydrodynamics as described in
\citet{Keller14}.
Haloes are selected at redshift $z=0.0$ from parent dissipationless
simulations of size 60, 20 and 15 $h^{-1}$Mpc, presented in
\citet{Dutton14}, which adopt a flat $\Lambda$CDM cosmology with
parameters from the \citet{Planck14}: Hubble parameter $H_0$= 67.1
\kms Mpc$^{-1}$, matter density $\Omegam=0.3175$, dark energy density
$\Omega_{\Lambda}=1-\Omegam=0.6825$, baryon density $\Omegab=0.0490$,
power spectrum normalization $\sigma_8 = 0.8344$ and power spectrum slope
$n=0.9624$.  Haloes are selected uniformly in log halo mass from $\sim
10$ to $\sim 12$ {\it without} reference to the halo merger history,
concentration or spin parameter.  Star formation and feedback is
implemented as described in \citet{Stinson06, Stinson13}.  Mass and
force softening are chosen to resolve the mass profile at $\lta 1\%$
the virial radius, which results in $\sim 10^6$ dark matter particles
inside the virial radius of all haloes at $z=0$. The motivation of
this choice is to ensure that the simulations resolve the galaxy
dynamics on the scale of the half-light radii, which are typically
$\sim 1.5\%$ of the virial radius \citep{Kravtsov13}.

Each hydro simulation has a corresponding dark matter only (DMO)
simulation of the same resolution.  In some cases these simulations
are in different evolutionary states, due to a major merger occurring
at a different time. In order for a fair comparison between hydro and
DMO simulations, we remove four haloes for which there is a large
difference ($>$ factor 1.3) between the halo masses in the hydro and
DMO simulations \citep[eee fig.1 in][]{Dutton16}.  In addition we
remove the three most massive haloes (g1.77e12, g1.92e12, g2.79e12),
as there is evidence that these have formed too many stars, in
particular near the galaxy centres.  The final sample consists of 83
simulations.

%% FIGURE 1
\begin{figure*}
\includegraphics[width=0.45\textwidth]{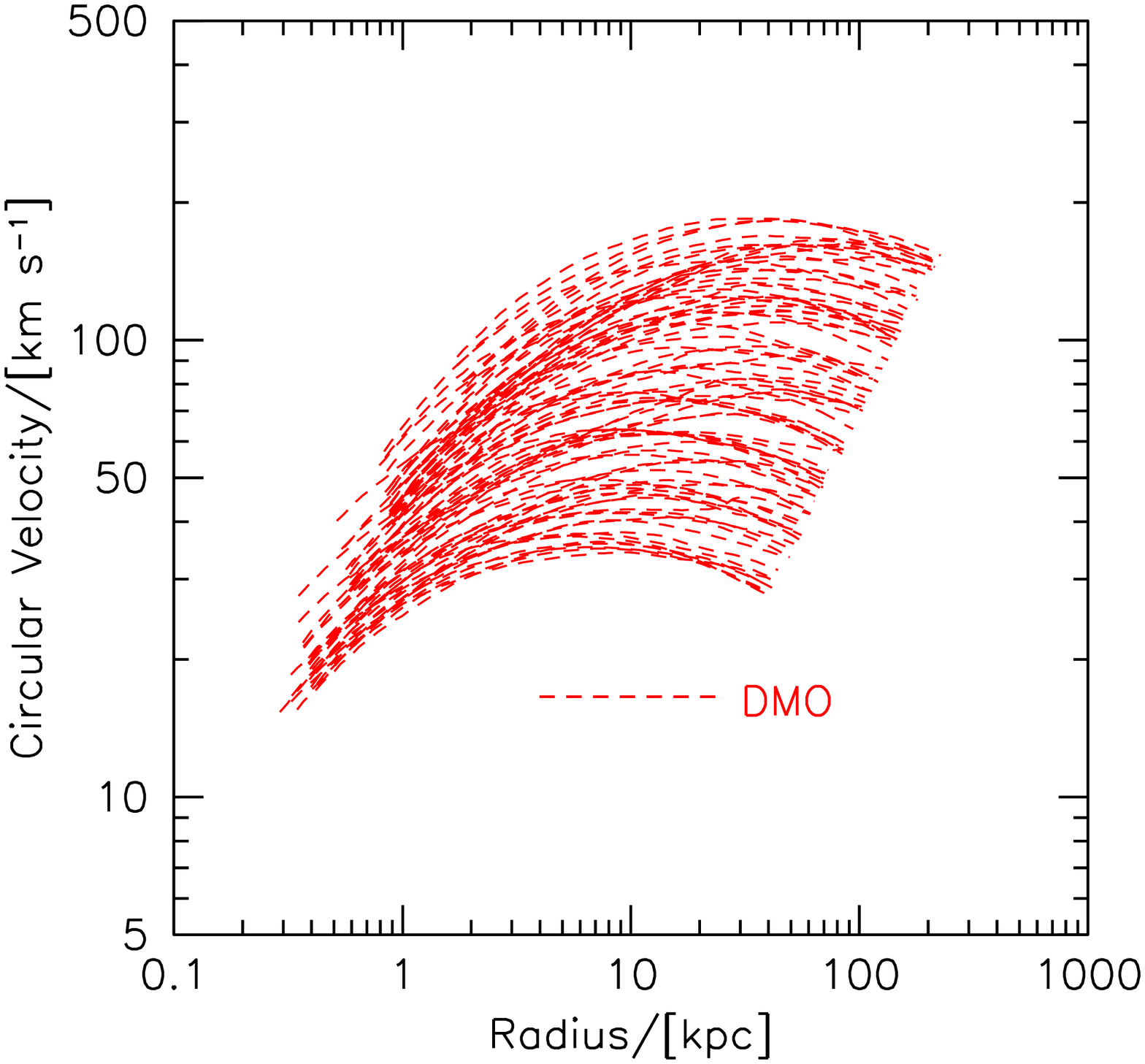}
\includegraphics[width=0.45\textwidth]{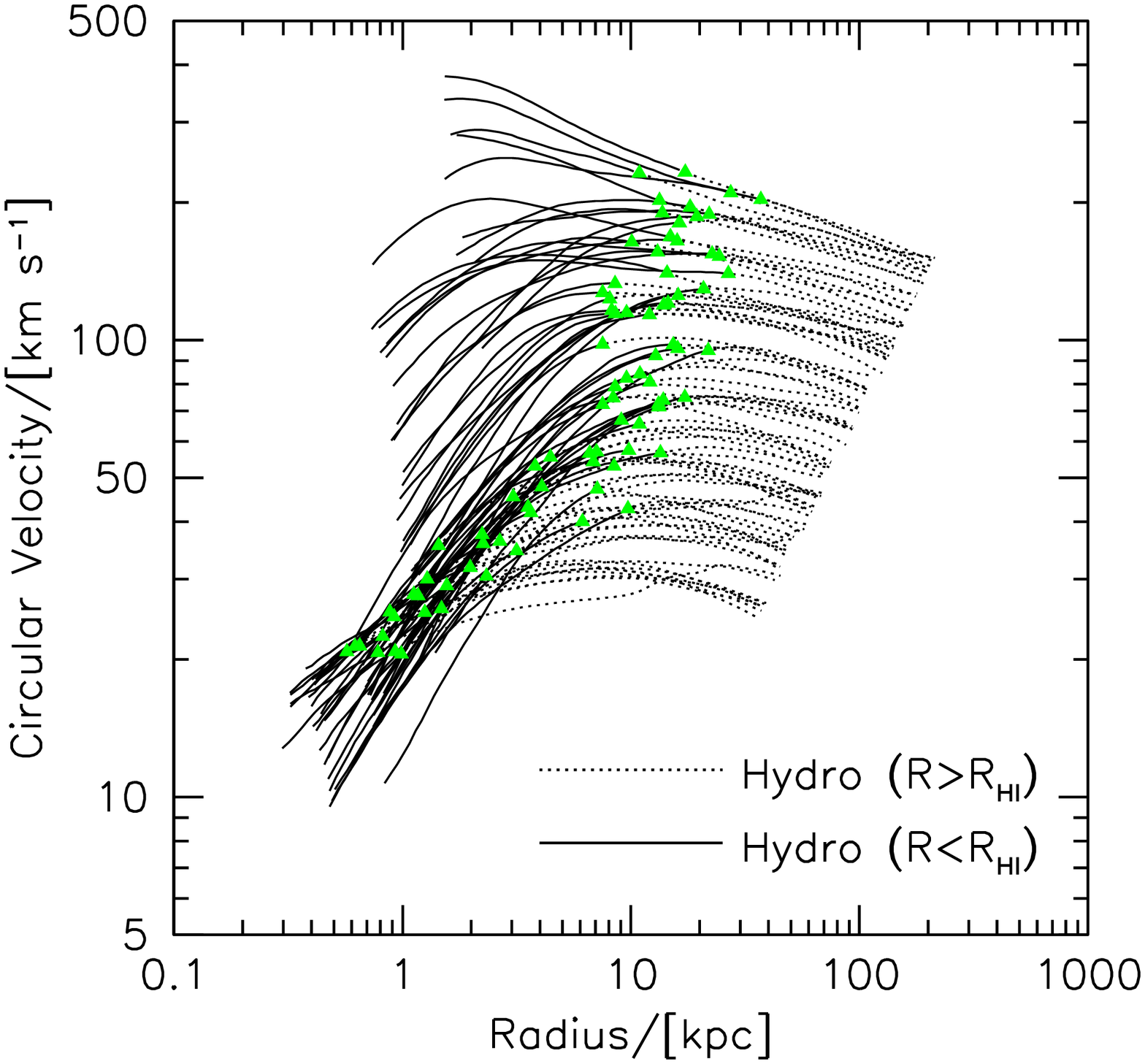}
\caption{Circular velocity profiles for the primary simulated galaxies
  used in this paper. Lines are plotted out to the virial radius.
  Black lines show circular velocity for hydrodynamical simulations,
  solid lines for radii smaller than the \HI\, radius, dashed for larger
  radii,  red dashed lines show circular velocity for the dark matter
  only (DMO) simulations. The circular velocity at the \HI\, radius is
  shown with a green triangle. There is more variety in the profiles
  from the hydro simulations at radii below $\sim 0.1 R_{200}$. }
\label{fig:rotcurve}
\end{figure*}

\subsection{Derived galaxy and halo parameters}

Haloes in NIHAO zoom-in simulations were identified using the
MPI+OpenMP hybrid halo finder {\sc
  ahf}\footnote{http://popia.ft.uam.es/AMIGA} \citep{Gill04,
  Knollmann09}. {\sc ahf} locates local over-densities in an
adaptively smoothed density field as prospective halo centres. The
virial masses of the haloes are defined as the masses within a sphere
whose average density is 200 times the cosmic critical matter density,
$\rhocrit=3H_0^2/8\pi G$.  The virial mass, size and circular velocity
of the hydro simulations are denoted: $M_{200}, R_{200}, V_{200}$.
The corresponding properties for the DMO simulations are
denoted with a superscript, ${\rm DMO}$.  For the baryons, we calculate
masses enclosed within spheres of radius $r_{\rm gal}=0.2R_{200}$,
which corresponds to $\sim 10$ to $\sim 50$ kpc.  The stellar mass
inside $r_{\rm gal}$ is $M_{\rm star}$, the neutral gas inside $r_{\rm
  gal}$ is $M_{\rm neut}\equiv 1.33 M_{\rm HI}$, where the neutral
hydrogen, \HI, mass is computed following \citet{Rahmati13} as
described in \citet{Gutcke17}.  The galaxy baryonic mass is defined as
$M_{\rm gal}=M_{\rm star}+M_{\rm neut}$, while the virial baryonic
mass $M_{\rm bar}$ is all of the stars and gas within $R_{200}$.  We
measure the circular velocity at a number of radii as discussed below.

%% FIGURE 2
\begin{figure}
\centerline{
  \includegraphics[width=0.45\textwidth]{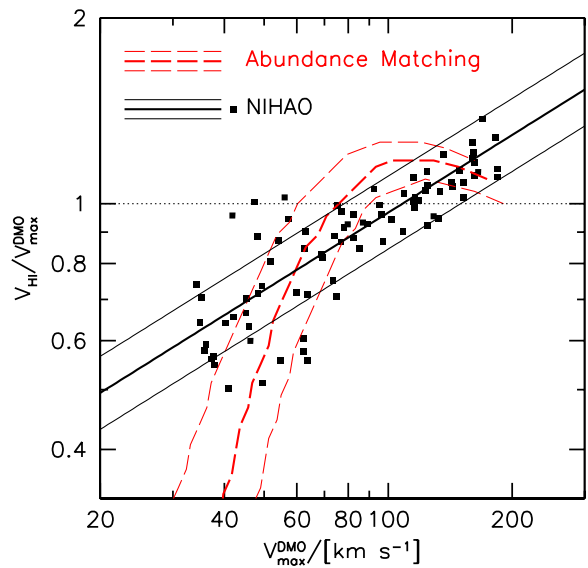}
}
\caption{Relation between circular velocity at the \HI\, radius, $\VHI$,
  and the maximum circular velocity of the DMO simulation,
  \Vmaxdmo. Points show NIHAO simulations, black lines show a power-law fit with $1\sigma$
  scatter, red lines show results from halo abundance matching
  \citep{Papastergis15}.  }
\label{fig:vv}
\end{figure}

\subsection{Definitions of Circular Velocity}
Since the circular velocity profiles of galaxies are not constant, the
choice of circular velocity one adopts will impact the slope,
zero-point and scatter of the TF relation \citep[e.g.,][]{Brook16,
  Bradford16}.  Depending on the goals of the study, different
definitions should be adopted. For example, measurements at small
radii are more sensitive to the distribution of the baryons and the
halo response, while measurements at large radii are a better probe of
the halo mass.

%% FIGURE 3
\begin{figure*}
  \centerline{
  \includegraphics[width=0.85\textwidth]{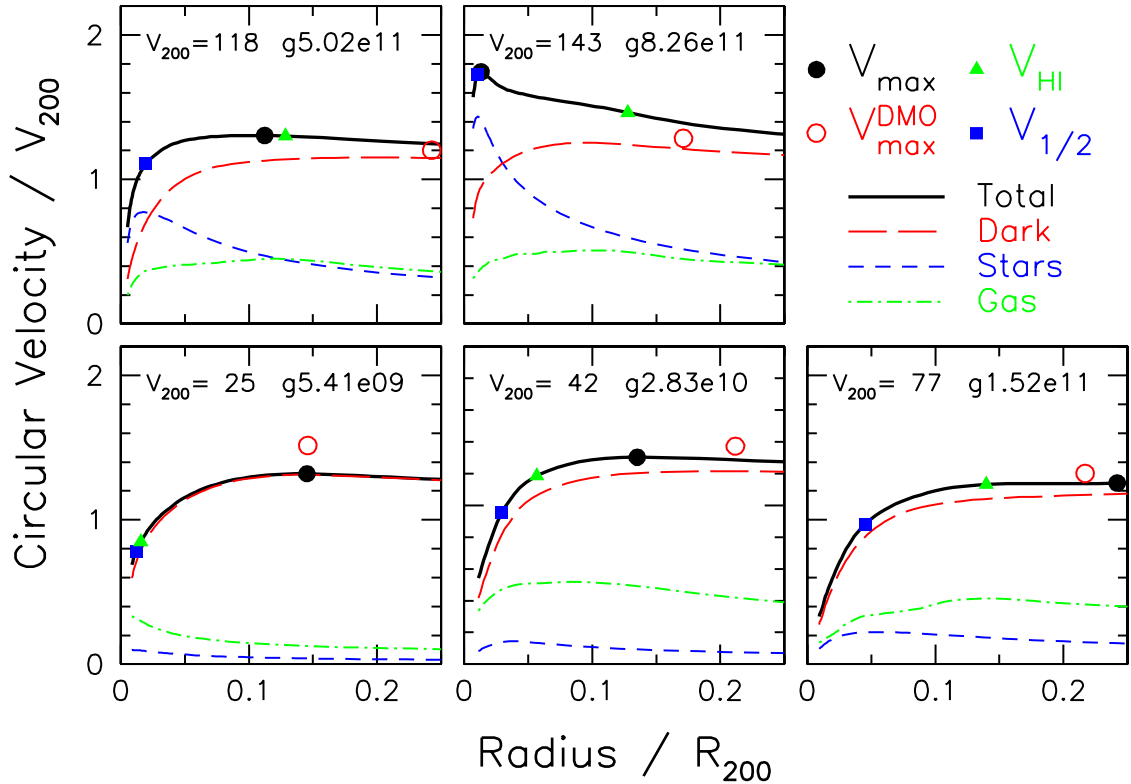}    
}
\caption{Example circular velocity profiles ordered by halo virial
  velocity, $V_{200}$.  The various lines correspond to the total
  (black solid), dark matter (red long-dashed), stars (blue short
  dashed), gas (green dash-dotted) components. Symbols show a variety of
  characteristic circular velocities $V_{1/2}$ (measured at the radius
  enclosing 50\% of the stellar mass, blue square), $V_{\rm HI}$
  (measured at the radius enclosing 90\% of the \HI\, gas, green
  triangle), $V_{\rm max}$ (maximum circular velocity of the hydro
  simulation, black circle) and $V_{\rm max}^{\rm DMO}$ (maximum
  circular velocity of the DMO simulation, red open circle). The \HI\,
  radius typically occurs in the flat part of the rotation curve,
  except for the lowest mass galaxies.}
\label{fig:rotcurve2}
\end{figure*}

Fig.~\ref{fig:rotcurve} shows circular velocity profiles,  $V_{\rm
  circ}=\sqrt{G M(r)/r}$, for the primary galaxy in each zoom-in
simulation. It shows the diversity of circular velocity profiles of
the hydro simulations (right panel, black lines), compared to the
dissipationless simulations (left panel, red dashed lines). In
particular, at small radii, both lower and higher velocities (relative
to the dissipationless simulations) are seen.  The lower velocities
are due to the expansion of the dark matter halo
\citep{Dutton16,Tollet16}, while the higher velocities are primarily
due to the collapse of baryons.  The green triangles show the circular
velocity at the \HI\, radius, $R_{\rm HI}$, which is defined to enclose
90\% of the \HI\, flux in a face-on projection.  This typically occurs
at $\sim10\%$ of the virial radius, except for the lowest mass
galaxies where the \HI\, radius occurs at a smaller fraction of the
virial radius due to a significant amount of the cold gas $T \lta
30000$K being ionized.

The \HI\, velocity typically traces the maximum circular velocity of the
DMO simulation (Fig.~\ref{fig:vv}), again with the exception of the
low mass galaxies. The relation between \VHI and \Vmaxdmo is well fitted
with a power law with a slope steeper than unity:
\begin{equation}
\log_{10}\left(\frac{V_{\rm HI}}{[\kms]}\right)=1.83 +1.42\left[\log_{10}\left(\frac{V_{\rm max}^{\rm DMO}}{[\kms]}\right) - 1.89\right]
\end{equation}
with a scatter of 0.06 dex in \VHI at fixed \Vmaxdmo.  For comparison,
the red lines show the relation between \VHI and \Vmaxdmo applying halo
abundance matching to the velocity function \citep{Papastergis15},
which encouragingly broadly agrees with the NIHAO simulations.  There
are some caveats in a direct comparison \citep[see][for a more
  detailed discussion]{Maccio16}. Briefly, the observed velocity
function is based on \HI\, linewidths, which does not necessarily trace
the circular velocity at the \HI\, radius. An inclination correction is
applied in \citet{Papastergis15} assuming thin rotating discs,
whereas, there can be non-negligible pressure support especially in
low mass galaxies. The abundance matching assumes no scatter in galaxy
velocity at fixed halo velocity, whereas simulations find a large
scatter.

Fig.~\ref{fig:rotcurve2} shows a representative sample of five
circular velocity curves for haloes with virial velocity ranging from
$V_{200}=25\kms$ to $V_{200}=143\kms$. The thick black lines show the
total circular velocity, which is broken down into the contributions
from stars (blue short-dashed lines),  dark matter (red long-dashed
lines) and gas (green dash-dotted lines). Low mass haloes are dark
matter dominated at all radii.  The baryons and in particular the
stars make a larger contribution in more massive haloes.  In haloes
with $V_{200}\gta 100 \kms$ the baryons dominate the centre.

%% FIGURE 4
\begin{figure*}
\centerline{
\includegraphics[width=0.85\textwidth]{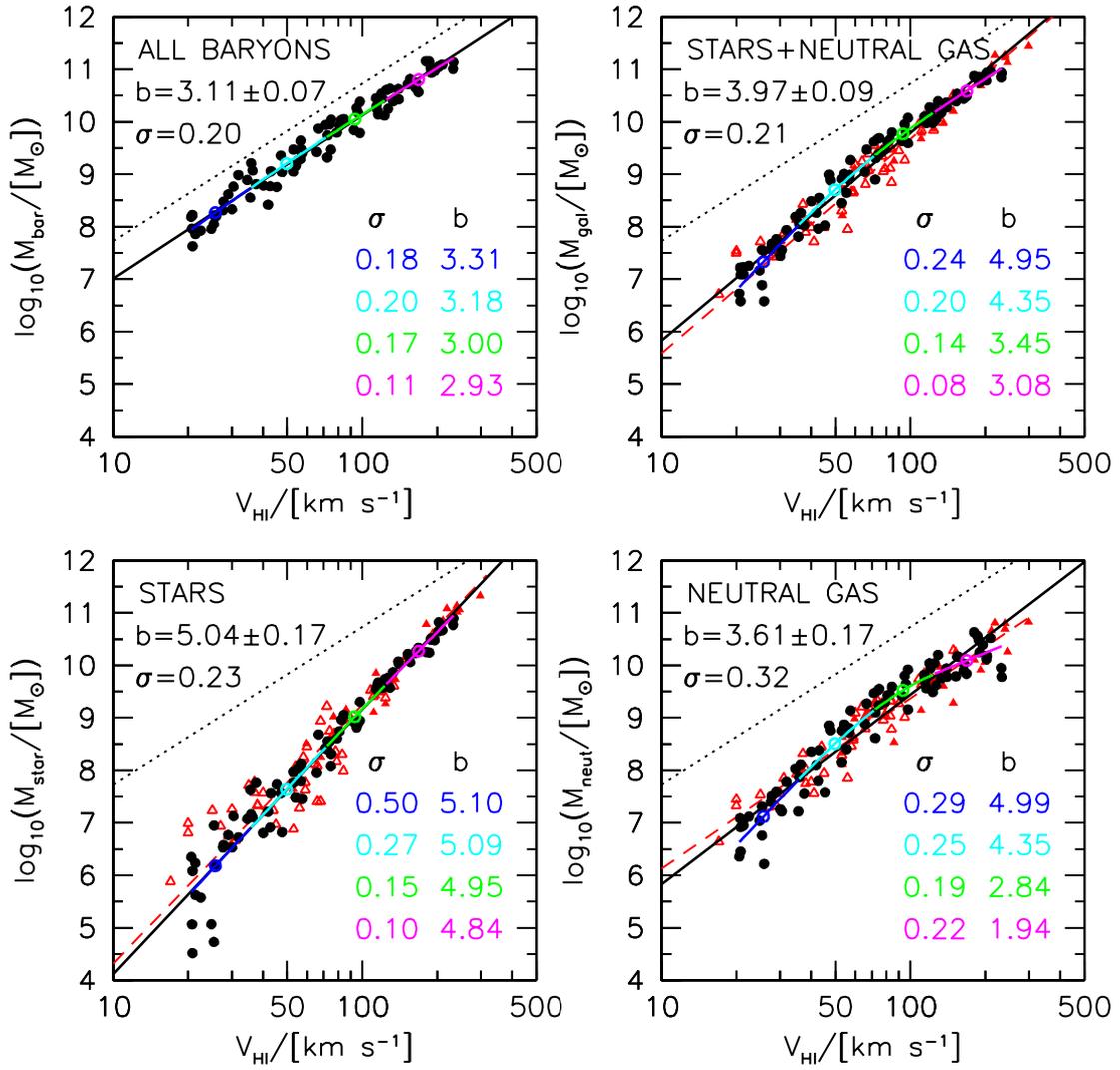}
}
\caption{TF relations for velocities measured at the
  \HI\, radius (in NIHAO simulations, black filled circles), and in the
  `flat' part of the rotation curve for observations (red
  triangles). Upper left: total baryonic mass within the virial
  radius, $M_{\rm bar}$; upper right: stellar+neutral gas, $M_{\rm
    gal}$; lower left: stellar mass, $M_{\rm star}$; lower right:
  neutral gas, $M_{\rm gas}$. Observations are shown  from
  \citet[][open triangles]{McGaugh12} and \citet[][filled
    triangles]{McGaugh15}.  For reference, the dotted line shows the
  virial relation: $M_{\rm b}=\Omegab/\Omegam V^3/hG
  $. The numbers in the upper left corner of each panel give the
  slope, $b$, and scatter, $\sigma$, for a global straight line fit
  (shown with solid black line). The lower right corner gives the
  slope and scatter for fits in quartiles of velocity (shown as
  colored lines).}
\label{fig:mass_vflat}
\end{figure*}

For each galaxy circular velocities at four radii are shown: $V_{1/2}$
(blue  square) measured at the radius enclosing (within a sphere) half
the stellar mass; $\VHI$ (green filled triangle) measured at the \HI\,
radius; $V_{\rm max}$ (black circle) the maximum circular velocity of
the hydro simulation, and $\Vmaxdmo$ (red open circle)
the maximum circular velocity of the dark matter only simulation.  The
stars are confined to a few percent of the virial radius, and thus the
$V_{1/2}$ typically traces the rising part of the velocity profile.
The \HI\, radius typically occurs at 5-15 percent of the virial radius,
and thus traces the `flat' part of the rotation curve. \VHI is
typically 20 to 40 percent higher than the virial velocity, $V_{200}$.
 The maximum circular velocity of the DMO simulation occurs at
$\sim 20$ percent of the virial radius, and thus can often be approximated
by $V_{\rm HI}$.

%% FIGURE 5
\begin{figure*}
\centerline{
\includegraphics[width=0.85\textwidth]{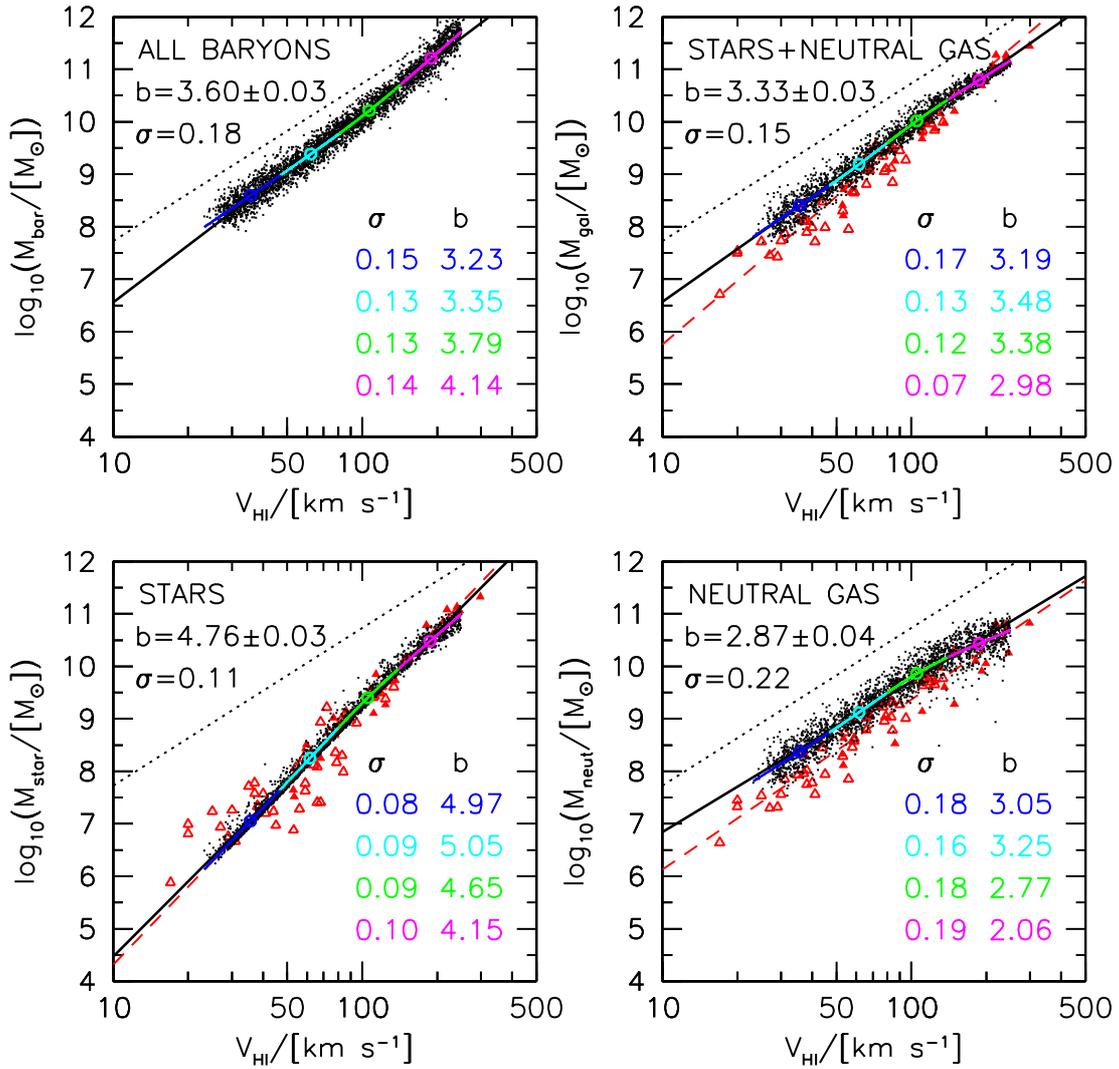}
}
\caption{As Fig.~\ref{fig:mass_vflat} but using the SAM (black dots) of \citet{Dutton12}.}
\label{fig:mass_vflat_sam}
\end{figure*}

%%%%%%%%%%%%%%%%%%%%%%%%%%%%%%%%%%%%%%%%%%%%%%%%%%%%%%%%%%%%%%%%%%%%%%%%%%%%
%% SECTION 3 MASS VELOCITY RELATIONS
%%%%%%%%%%%%%%%%%%%%%%%%%%%%%%%%%%%%%%%%%%%%%%%%%%%%%%%%%%%%%%%%%%%%%%%%%% %%%%%
\section{Tully Fisher relations}
\label{sec:results}
We now construct various TF relations. We fit these relations with a
straight line $y=a + b (x-x_0)$ using linear least squares. Here $x_0$
is the mean of $x$. In general, the independent variable is the
logarithm of a velocity $x=\log_{10}(V/[\kms])$, while the dependent
variable is the logarithm of a mass $y=\log_{10}(M/[\Msun])$.  We
estimate uncertainties on $a$ and $b$ using jackknife re-sampling. We
calculate the scatter about the best-fitting relation using the 68th
percentile of the absolute residuals, $p_{68}$, which is less
sensitive than the standard deviation to outlier points.

To gauge the dependence of the slopes and scatter on velocity, we
divide the velocities into quartiles and fit each quartile
independently. This yields a mean value of $y$ in four velocity bins,
which we spline interpolate to create a mean relation across the full
$x$ range. This ensures continuity in the mean
relations across the bin boundaries. We then calculate the scatter
about this mean relation, and the local slopes in the four velocity
bins.  The slopes and scatter in each velocity quartile are shown in
the lower right corner of Fig.~\ref{fig:mass_vflat}.
 
\subsection{Comparison with observed TF relations}
\label{sec:vm}

We start with a comparison with observations, and thus use the
observationally accessible $\VHI$. In Fig.~\ref{fig:mass_vflat},
black filled circles  are the NIHAO simulations, while red open and
solid triangles show observations from \citet{McGaugh12} and
\citet{McGaugh15}, respectively. For the data from \citet{McGaugh15}
stellar masses are obtained from $3.6\mu$m photometry assuming a
stellar mass-to-light ratio of 0.45. For the observations, $\VHI$ is
measured in the `flat' part of the \HI\, rotation curve. Recall that in the
simulations, it is measured at a radius enclosing 90\% of the neutral gas.
A potentially important caveat to our comparison is the observations
are using rotation velocities, while in the simulations, we are using
(spherically symmetric) circular velocity. 

As we will see, in general, there is good agreement between the
simulations and observations, although in detail there are
differences.  We discuss each of the TF relations in turn
below.  In summary, the simulations have plausible amounts of stars and
neutral gas over a wide range of halo masses, and thus can be used to
inform us about the stochasticity of galaxy formation in a $\LCDM$
universe.

As an additional reference point for the expectations for these
scaling relations in a \LCDM universe, we show the results from the
SAM of \citet{Dutton12}, updated here using the {\it Planck} cosmology, in
Fig.~\ref{fig:mass_vflat_sam}. The SAM and NIHAO simulations give
similar results. For example, the stellar TF relation is steepest,
while gas TF relation is the shallowest.

{\it Stellar TF relation:} there is a good agreement between
observations and NIHAO simulations of the stellar TF relation (lower
left panel). Simulations have a slope of 5.04$\pm0.17$, while
observations have 4.93$\pm 0.17$.  The total scatter in the
simulations is 0.23 dex, and the intrinsic scatter in the observations
is 0.15 dex \citep{McGaugh15}. However, the scatter is clearly mass
dependent both in simulations and observations, with higher scatter at
lower velocities. One might guess a physical cause of this trend is
that star formation is a stochastic process, and thus when more stars
are formed the variations average out. However, when compared to the
halo velocity TF relations (see \S{\ref{sec:halotf}}), we see the
scatter is almost independent of mass, except for the very lowest mass
haloes. Thus the variation in scatter is due to how the baryons
influence $\VHI$. The smaller scatter at high $\VHI$ is due to the
baryons contributing to $\VHI$ (i.e., more baryons give higher
velocity), while the larger scatter at small $\VHI$ is due to the
variation in $\VHI/V_{200}$, presumably due to $\VHI$ tracing the
rising part of the rotation curve, and thus being more sensitive to
the exact radius the rotation is measured at.

{\it Gas TF relation:} the lower right panel shows the relation between
neutral gas mass and \HI\, circular velocity. The slopes of the
simulations and observations are $3.61\pm0.17$ and $3.24\pm0.28$. The
scatter is roughly independent of velocity at $\sim 0.25$ dex in both
simulations and observations. Note that the SAM appears to
over-predict the neutral gas mass. However, the SAM does not take into
account the effects of ionization on the gas mass, which is
substantial in the hydro simulations.  

{\it Baryonic TF relation:} the upper panels show the baryonic TF
relations. For the simulations, the upper right panel shows the stars plus
neutral gas within $0.2R_{200}$ (i.e., the observable baryons), while
the upper left panel shows all the baryonic mass (i.e., stars plus gas)
within $R_{200}$.  The observations are shown only in the upper right
panel and consist of stars plus neutral gas (atomic + molecular).
The observed slope is $4.09\pm0.15$ obtained by us using the same
fitting technique we apply to the simulations. Reassuringly, this
slope is consistent with fits from the original authors
\citep{McGaugh12, McGaugh15}.  For the simulations the `observable'
baryons have a slope of $3.97\pm0.09$, in excellent agreement.  It
should be noted that the slope of the observed BTF strongly depends on
the stellar mass-to-light ratio with values ranging from 3.4 to 4.0
for plausible M/L \citep{Lelli16}. Here we adopt the calibrations of
\citet{McGaugh12} and \citet{McGaugh15}, which are on the high side of the
plausible range and hence give steeper slopes.  The full baryonic mass
has slope of $3.11\pm0.07$, very close to the nominal $\LCDM$ value of
3. If one assumes $\VHI \propto V_{200}$ thus implies a constant
baryon mass fraction.  However, as we see below, this assumption is
not valid in our simulations. The actual baryon fractions are lower in
lower mass haloes.

%% FIGURE 6
\begin{figure}
\centerline{
  \includegraphics[width=0.45\textwidth]{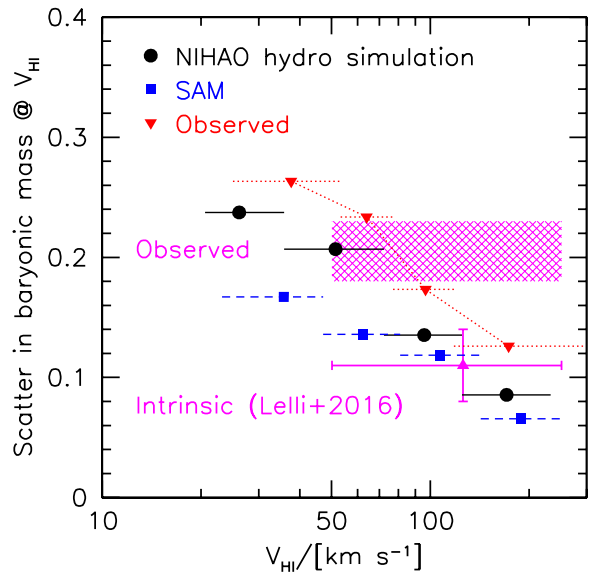}
}
  \caption{Scatter of the baryonic TF relation $M_{\rm gal}$ versus
  $\VHI$ as a function of velocity. Black filled circles show results
  from the NIHAO simulations, while the blue squares  dashed lines
  show results from the SAM used in \citet{Dutton12}. Horizontal error
  bars correspond to the range in each velocity bin. Red triangles
  show the observed scatter using observations from \citet{McGaugh12}
  and \citet{McGaugh15}. The observed and intrinsic scatter from
  \citet{Lelli16} are shown with the magenta shaded region and error
  bar, respectively.}
\label{fig:btf_scatter}
\end{figure}

%% FIGURE 7
\begin{figure*}
\centerline{
\includegraphics[width=0.4\textwidth]{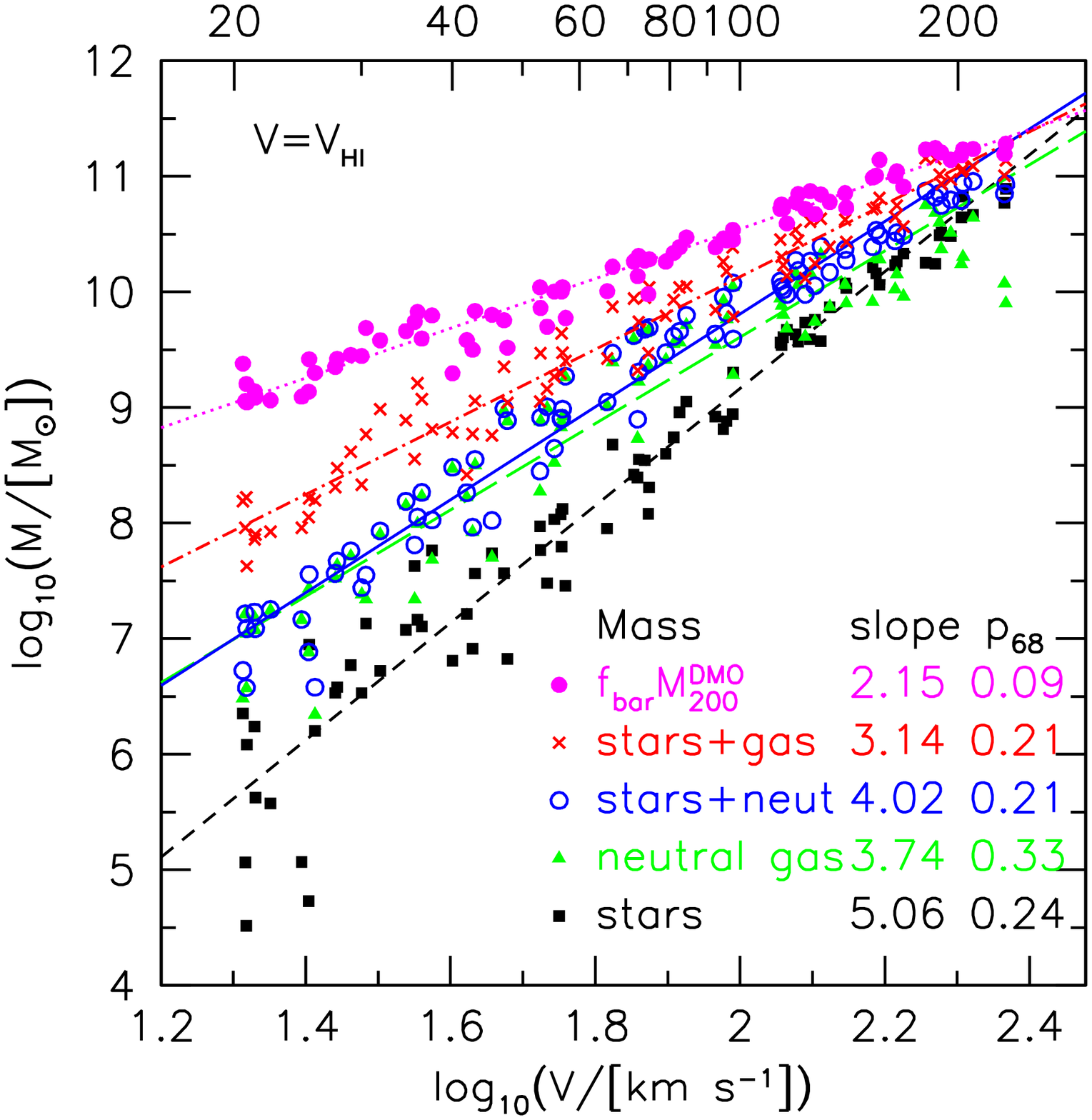}
\includegraphics[width=0.4\textwidth]{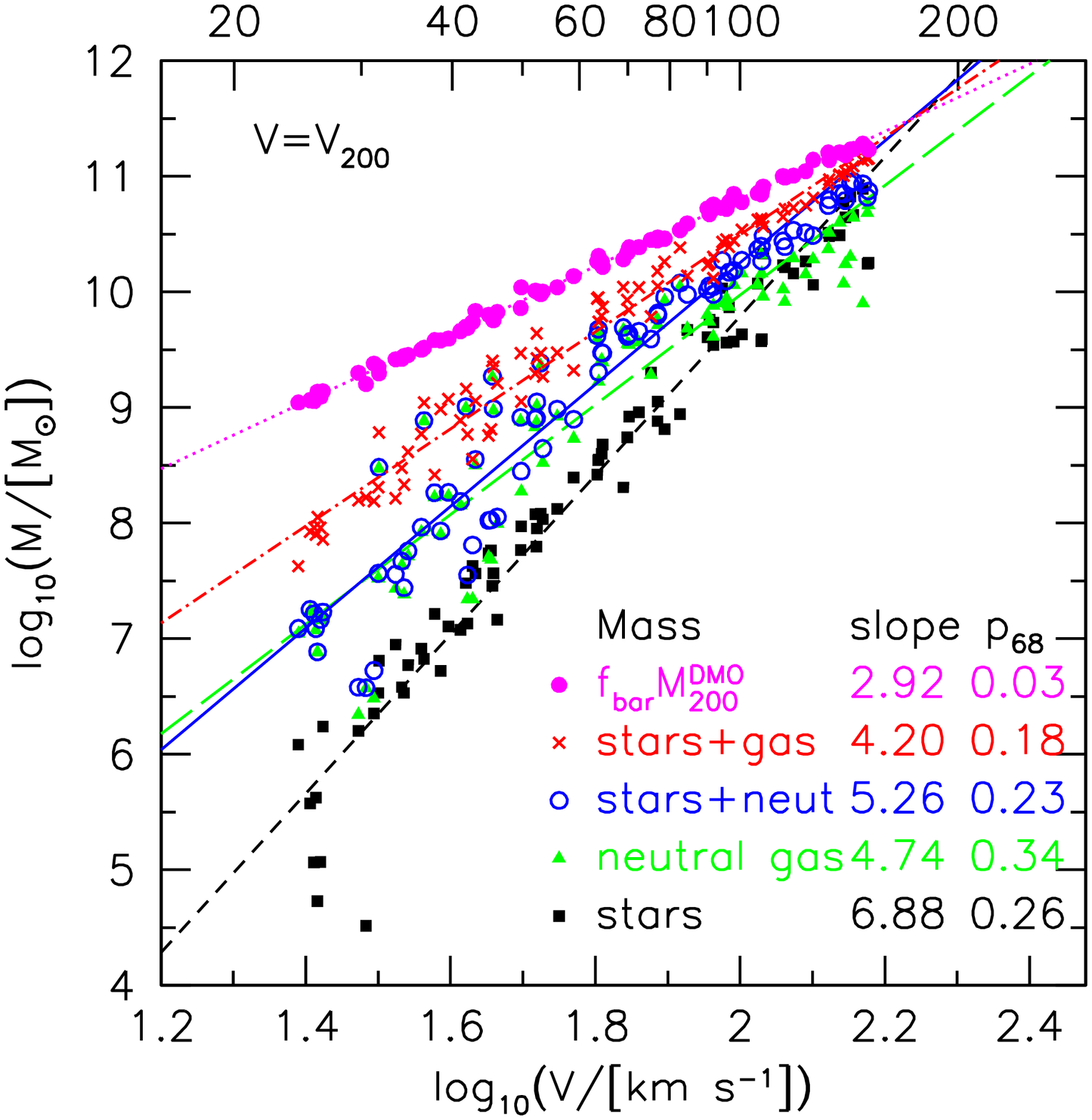}
}
\centerline{
\includegraphics[width=0.4\textwidth]{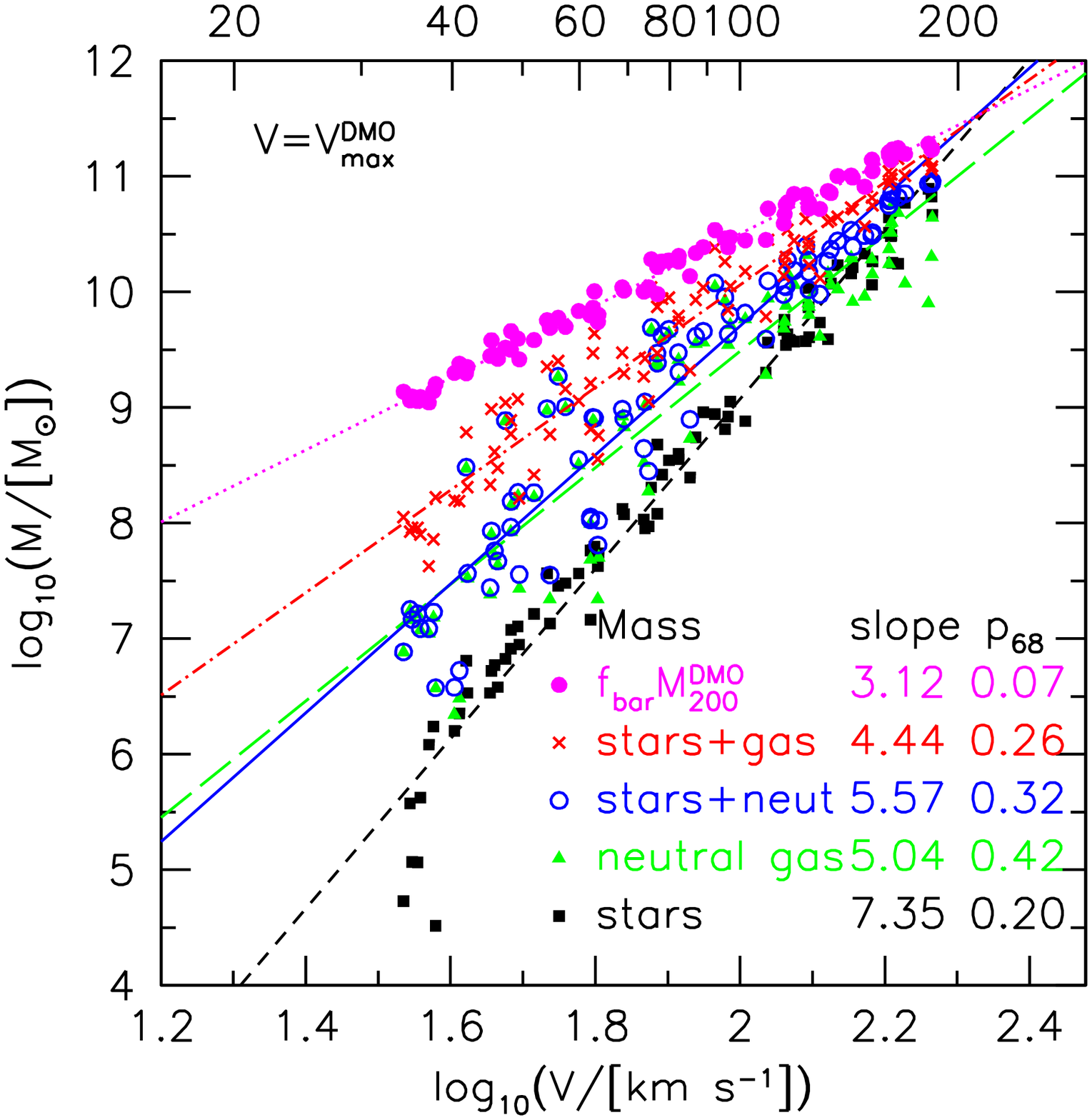}
\includegraphics[width=0.4\textwidth]{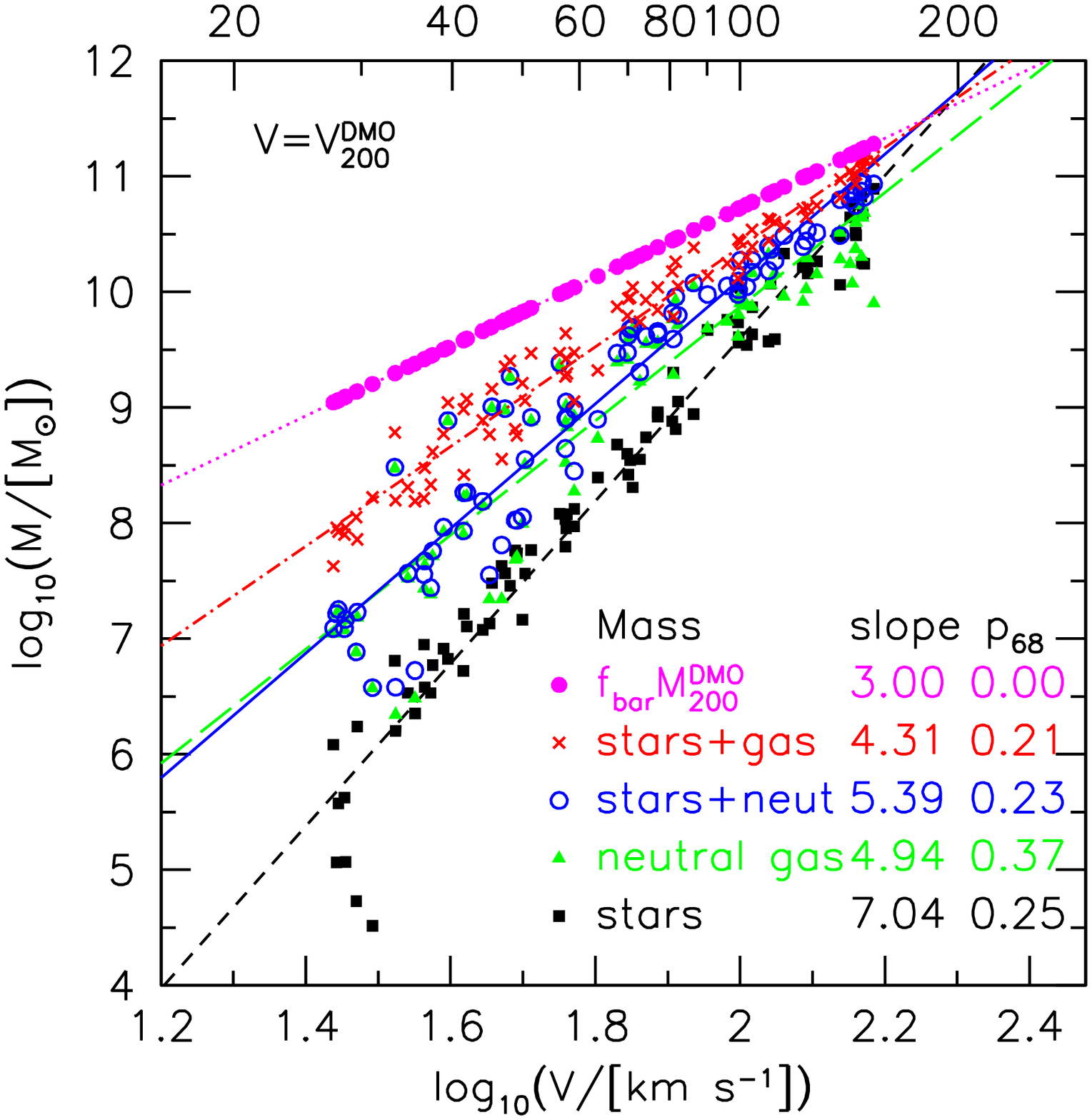}
}
\caption{TF relations for a variety of masses and velocities. The
  masses are: stars (black squares), \HI\, gas (green triangles), stars+
  \HI\, gas (blue open circles), stars + all gas (red crosses) and the
  total baryons associated with the DMO simulation,
  $(\Omegab/\Omegam)M^{\rm DMO}_{200}$ (magenta filled circles). The
  velocities are: \VHI (upper left), $V_{200}$ (upper right), $V_{\rm
    max}^{\rm DMO}$ (lower left), $V_{\rm 200}^{\rm DMO}$ (lower
  right). Each relation is fitted with a power-law, whose slope and
  scatter are given in the lower right corner. The data points for
  this figure are given in Table~\ref{tab:mvall}.}
\label{fig:mvall}
\end{figure*}

For galaxies with velocities in the range $50 \lta \VHI \lta
250 \kms$, the observed scatter and intrinsic scatter in the BTF is $\simeq
0.18-0.23$ and $\simeq 0.11\pm0.03$ dex, respectively
\citep{Lelli16}. Over the full velocity range probed by the NIHAO
simulations $20 \lta \VHI \lta 240 \kms$ the total scatter about a
power-law fit is 0.21 dex (Fig.~\ref{fig:mass_vflat}).  This is
comparable to the observed scatter, but much larger than the estimated
intrinsic scatter, nominally implying an inconsistency between our
$\LCDM$ simulations and observations.

However, in our simulations the scatter is strongly velocity
dependent,  with smaller scatter for higher velocity
galaxies. Fig.~\ref{fig:btf_scatter} shows the scatter of the NIHAO
simulations in quartile bins in velocity. Low velocity galaxies $20
\lta \VHI \lta 70 \kms$ have scatter of $\simeq 0.20 - 0.24$ dex,
while high velocity galaxies $70 \lta  \VHI \lta 240 \kms$ have
scatter of 0.08 to 0.14 dex in good agreement with the observed
intrinsic scatter on these scales.  Fig.~\ref{fig:btf_scatter} also
shows the SAM of \citet{Dutton12}, which has a scatter
that varies from 0.17 dex at low velocity to 0.07 dex at high
velocity.  The hydro simulations, which include more physical
processes than the SAM, would nominally be expected to
have larger scatter, which is indeed the case.

Given the tentative discrepancy we find between theory and observation
at low velocities ($\VHI \lta 70 \kms$) it will be interesting to
study them further, both to refine the predictions and obtain larger
observational samples.  An important consideration is that the
rotation curves rise more slowly than in high mass galaxies (see
Fig.~\ref{fig:rotcurve2}).  This means that the observed neutral
hydrogen does not always reach the flat part of the rotation curve. It
is debated whether or not this introduces a bias in the BTF at low
masses \citep{Brook16, Lelli16}, and is thus an important topic for
future studies to resolve.

%% FIGURE 8
\begin{figure*}
  \centerline{  
    \includegraphics[width=0.9\textwidth]{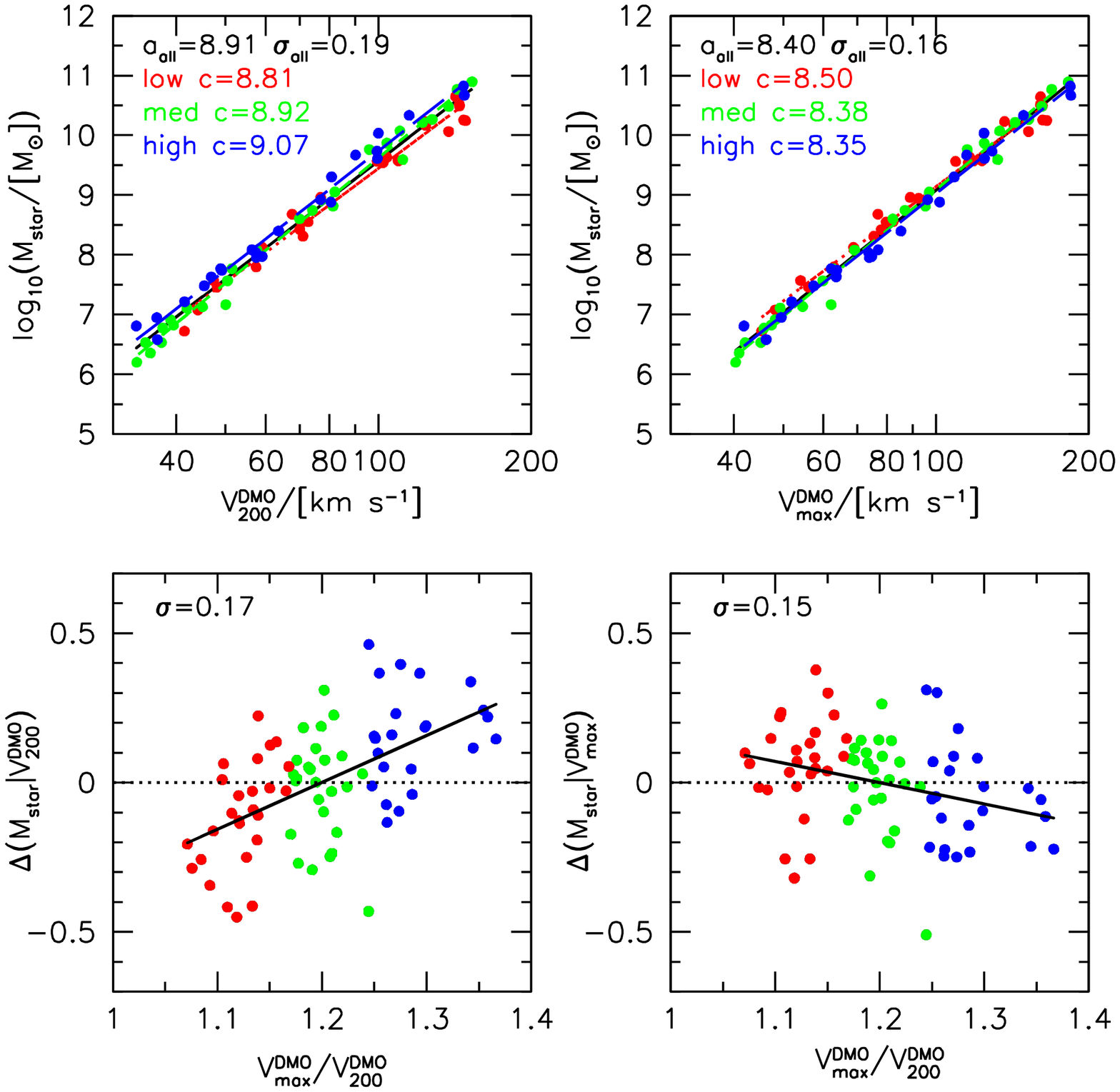}
    }
  \caption{Dependence of scatter in stellar mass - halo velocity
    relation on halo concentration. The upper panels show the mass -
    velocity relations, using virial velocity, $V_{200}^{\rm DMO}$
    (left), and maximum circular velocity, $V_{\rm max}^{\rm DMO}$
    (right) from the DMO simulations. The zero-point, $a_{\rm all}$,
    and scatter, $\sigma_{\rm all}$ from a linear fit (solid black
    lines) are shown at the top of each panel. The subsequent numbers
    show the zero points for fits to low, average and high
    concentration haloes.  The lower panels show the residuals about
    the best fit  relation versus $V_{\rm max}^{\rm DMO}/V_{200}^{\rm
      DMO}$ which is our proxy for halo concentration. In all panels
    points are colour coded by concentration, with red for low, green
    for average and blue for high.  At fixed virial velocity  higher
    concentration haloes are offset to higher stellar masses, while at
    fixed maximum circular velocity higher concentration haloes are
    offset to lower stellar masses.  The scatter about the mass
    residual versus concentration relation is given in the upper left
    corner, and shows that including concentration reduces the scatter
    in the mass - velocity relation. }
\label{fig:vm_c}
\end{figure*}

\subsection{Dark halo velocity versus mass relations}
\label{sec:halotf}

Fig.~\ref{fig:mvall} shows the relations between various galaxy masses
(total baryonic mass, galactic baryonic mass, stellar mass, and
neutral gas mass) and various velocities ($\VHI, V_{200}, V_{200}^{\rm
  DMO}, \Vmaxdmo$).  Recall that $V_{200}$ and $V_{200}^{\rm DMO}$ are by
definition equivalent to halo mass since $V_{200}=(GM_{200}h)^{1/3}$.
We fit each relation with a power-law over the full velocity range.
The best-fitting relations are shown with lines of the same colour as the
respective data points. The shallowest relation in each panel involves
the total available baryonic mass (magenta), which by definition has a
slope of 3 in the lower left panel.  All of the relations with galaxy
mass components have steeper slopes, corresponding to lower efficiency
of galaxy formation (whether it be defined as neutral gas, stars or
total baryons). 

The upper right panel of Fig.~\ref{fig:mvall} shows the relation
between mass and virial velocity of the hydro simulation,
$V_{200}$. All these relations are steeper than the corresponding
relations using $\VHI$.  Note also that lower mass haloes have
retained a smaller fraction of their baryons \citep[compare red and
  magenta points and lines; see also][]{Wang16}.  The scatter
increases as we go from the total baryonic mass (red crosses), to the
galactic baryonic mass (blue circles), to the stellar mass (black
squares)  and finally to the neutral gas mass (green triangles). This
supports the notion that the baryonic mass is a more fundamental
galaxy property than either stellar or gas mass.

The lower left panel shows results  using the maximum circular
velocity from the DMO simulation, $\Vmaxdmo$, while the lower right panel
uses the virial circular velocity from the DMO simulation,
$V_{200}^{\rm DMO}$.  The slopes are slightly steeper with
$V_{200}^{\rm DMO}$ than $V_{200}$ due to increased mass loss from
lower mass haloes, and steeper still using $\Vmaxdmo$ because lower mass
haloes have higher $\Vmaxdmo/V_{200}$ due to the higher average
concentrations (recall the concentration versus velocity relation for CDM
haloes goes like $c\propto M_{200}^{-0.1}\propto V_{200}^{-0.3}$).

Comparing the different panels in Fig.~\ref{fig:mvall}, we see that
stellar mass correlates better with maximum halo velocity than virial
velocity, while neutral gas mass and total baryonic mass correlates
better with virial velocity than maximum velocity. Thus, there is no
single velocity definition that minimizes the scatter in all baryonic
components.

\section{Implications for halo abundance matching}
\label{sec:ham}

The halo abundance matching technique is a powerful way to link the
masses of galaxies to the masses of dark matter haloes, under the
assumption of $\LCDM$ and cosmological parameters \citep{Conroy09}. It
can be used to understand how stars form over cosmic time.  In its
simplest form, the ansatz is that more massive galaxies live in more
massive dark matter haloes.  In our simulations scatter in stellar
mass at fixed halo virial or maximum circular velocity is small
$\simeq 0.20$ dex, and independent of halo velocity for $30 \lta
V_{200} \lta 160 \kms$ ($10^{10} \lta M_{200} \lta 10^{12.2}\Msun$).
This supports the halo abundance matching approach for central haloes.
At the lowest halo masses we probe $V_{200}\sim 25\kms$ ($M_{200}\sim
5 \times 10^9 \Msun$) the scatter starts to increase, consistent with
other simulation studies \citep[e.g.,][]{Sawala16}.

Using observations of the projected two-point galaxy clustering
\citet{Reddick13} showed that the peak circular velocity of the dark
matter halo (over the history of a halo) is more closely related to
stellar mass than the virial velocity of the halo.  Maximum circular
velocity (either at a given time, or over the history of a halo) is
often preferred over halo mass (or virial velocity) because (1) \Vmaxdmo
does not depend on the (somewhat arbitrary) definition of halo mass;
(2) \Vmaxdmo is less sensitive to the mass stripping that sub-haloes
experience and (3) \Vmaxdmo probes a smaller scale, which presumably has more
to do with galaxy formation than the virial scale.

In our galaxy formation simulations we indeed find that the maximum
circular velocity (at the present day) of the DMO simulation yields a
smaller global scatter (0.20 dex) in stellar mass than the virial
velocity (0.25 dex) of the hydro or DMO simulation
(Fig.~\ref{fig:mvall}). Removing the haloes with  $V_{200}^{\rm DMO} <
32 \kms$, which have very large scatter in stellar mass, results in
smaller scatters of 0.17 and 0.19 dex for $\Vmaxdmo$ and $V_{200}$,
respectively.

Going further, there has been recent discussion in the literature of a
dependence of the abundance matching on the concentration of the dark
matter halo \citep{Mao15, Paranjape15, Zentner16, Lehmann17}.  We thus investigate
further whether an additional parameter or a different definition of
halo velocity yields a tighter stellar mass - halo velocity relation.

\subsection{Dependence of stellar mass versus halo velocity relations on concentration}

The left panels of Fig.~\ref{fig:vm_c} show the stellar mass versus halo
virial velocity relation and its dependence on halo concentration.
Here as a proxy for concentration, we use the ratio between maximum and
virial circular velocities of the DMO simulation: $\Vmaxdmo/V_{200}^{\rm
  DMO}$.  At fixed halo velocity, $V_{200}^{\rm DMO}$, there is a
trend with concentration such that higher concentration haloes have
higher stellar masses.  This is plausibly because higher concentration
haloes, on average, form earlier \citep[e.g.,][]{Wechsler02}, and thus
have more time to form stars.  The overall scatter is 0.19 dex (upper
left), and can be reduced to 0.17 dex (lower left) by including the
correlation with concentration. A similar trend is seen in the stellar
mass versus halo mass relations of the {\sc eagle} simulations
\citep{Matthee17}. They found earlier forming and higher concentration
haloes have higher stellar mass.

The right panels of Fig.~\ref{fig:vm_c} show the same results by
using the maximum circular velocity of the halo in place of the virial
velocity. Now the correlation with concentration is weaker, and of the
opposite sign.  The change in sign of the correlation suggests that a
velocity measured somewhere between the virial radius and $0.2R_{200}$
(where $\Vmaxdmo$ occurs) would produce a tighter correlation with galaxy
stellar mass.   The overall scatter is 0.17 dex (upper right), and can
be reduced to 0.14 dex (lower right) by including the correlation with
concentration.

%% FIGURE 9
\begin{figure}
\centerline{
  \includegraphics[width=0.45\textwidth]{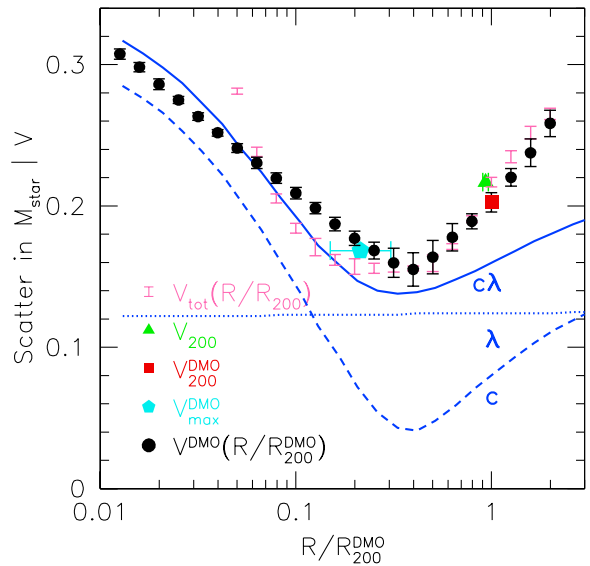} 
}
  \caption{Scatter of the stellar mass versus halo velocity relation for
    different definitions of  halo velocity.  Black circles show
    velocity  measured at various fractions of the virial radius in
    the DMO simulations. Error bars are from Jackknife
    re-sampling. Pink error bars show results using the total circular
    velocity at various radii in the hydro simulations.  The scatter
    is minimized for $R/R_{200}^{\rm DMO} \simeq 0.4$, providing a
    tighter relation than when using the maximum circular velocity of
    the DMO simulation (cyan pentagon) or virial velocity of the DMO
    (red square) or hydro (green triangle) simulation.  The solid blue
    lines show results from the SAM which includes scatter in
    concentration and spin ($c\lambda$). The SAM also has a minimum
    scatter at $R/R_{200}^{\rm DMO} \simeq 0.4$. Models run with just
    scatter in $c$ (dashed line), or $\lambda$ (dotted line) shows
    that concentration drives this minimum.}
\label{fig:vm_scatter2}
\end{figure}

\subsection{Dependence of scatter on halo velocity definition}

Fig.~\ref{fig:vm_scatter2} shows the scatter in the $\Mstar-V$
relation with velocity measured at various fractions of the virial
radius, $R/R_{200}$, in the DMO simulations.  We find that the minimum
scatter of $0.15$ dex occurs when the circular velocity is measured at
$\sim 0.4 R_{200}$. This corresponds to an overdensity of $\sim 1000
\rhocrit$.  This scatter is even smaller than when using the maximum
circular velocity of the DMO simulation (0.17 dex), which typically
occurs at $0.2 R_{200}$ (blue pentagon).  For reference the maximum
circular velocity occurs at $R_{V_{\rm max}} \simeq 2.2 R_{200}/c$.

\begin{figure}
\centerline{  
\includegraphics[width=0.45\textwidth]{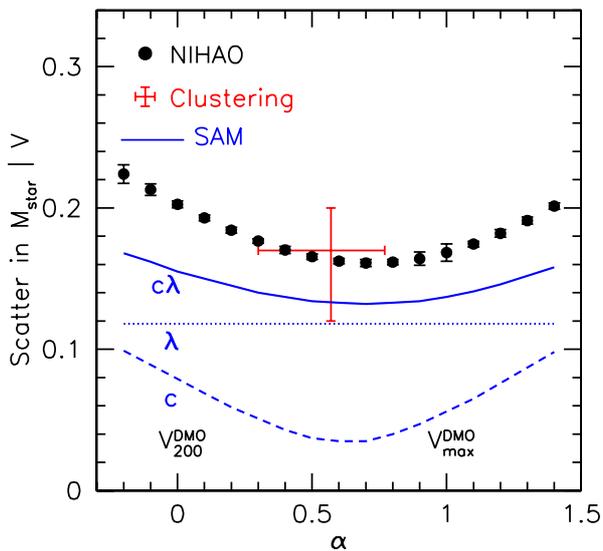}   
}
\caption{Scatter of the stellar mass versus halo velocity relation for
    $V_{\alpha}^{\rm DMO}=V_{200}(V_{\rm max}/V_{200})^{\alpha}$. The
    scatter is minimized for $\alphamin\simeq 0.7$, for both the NIHAO
    hydro simulations (black points) and the SAM (solid blue line),
    consistent with results from clustering \citep[][red error
      bar]{Lehmann17}. The dashed line shows a SAM that includes only
    scatter in halo concentration, while the dotted line shows a model
    that only includes scatter in halo spin. }
\label{fig:vm_scatter3}
\end{figure}

Observationally, \citet{Lehmann17} show that low redshift
clustering measurements from SDSS prefer a moderate amount of
concentration dependence (more than would be indicated by matching
galaxy luminosity to the peak halo mass, and less than would be
indicated by matching to the peak halo circular velocity).  Defining
the halo velocity as
\begin{equation}
V_{\alpha}=V_{200} (V_{\rm max}/V_{200})^{\alpha}
  \end{equation}
\citet{Lehmann17} found that the scatter in stellar mass was minimized at
$\sigma=0.17^{+0.03}_{-0.05}$ for $\alphamin=0.57^{+0.20}_{-0.27}$.  We
find similar results with our hydrodynamical
simulations. Fig.~\ref{fig:vm_scatter3} shows the scatter in the
stellar mass versus $V_{\alpha}^{\rm DMO}$ relation as a function of
$\alpha$. The scatter is minimized at $\alphamin\simeq 0.7$.

To help understand the origin of the results in
Figs.~\ref{fig:vm_c}-\ref{fig:vm_scatter3}, we show results from the SAM
of \citet{Dutton12}. Recall that there are just two parameters that
can cause variation in galaxy properties: halo concentration, $c\equiv
R_{\rm vir}/r_{-2}$, and halo spin, $\lambda\equiv J_{\rm vir}|E_{\rm
  vir}|^{1/2} / G M_{\rm vir}^{5/2}$. Note that in the SAM the halo
definition follows \citet{Bryan98}, while in the NIHAO simulations we
adopt an overdensity of $200 \rhocrit$, and thus for the analysis in
Figs.~\ref{fig:vm_c} and \ref{fig:vm_scatter3} we calculate $R_{200}$
and $V_{200}$. Both $c$ and $\lambda$ are independent log-normally
distributed at redshift $z=0$ with $\sigma_{\ln c}=0.25$ and
$\sigma_{\ln\lambda}=0.53$ \citep{Maccio08}.  For a given halo, the
spin is assumed to be constant with time. The scatter in the
concentration is correlated with the formation history of the halo
following \citet{Wechsler02}.

Remarkably, the SAM has a minimum scatter at the
same radii $R\sim 0.4 R_{200}$ and $\alphamin \sim 0.7$. The fact we find
similar results for the SAM as the full hydro simulation suggests that
there is a simple physical origin of these trends.  In addition to the
default model, we run models with only scatter in $c$ or
$\lambda$. These show that the minimum in scatter is driven solely by
the scatter in concentration parameter.

%% FIGURE 10

\subsection{Halo structure or formation time?}
As discussed in \citet{Zentner16} the halo concentration is correlated
with two physical properties of the dark matter halo that could affect
the star formation efficiency: the formation time of the halo and the
potential depth. Earlier forming haloes will have more time to form
stars \citep[e.g., see Fig.12 of][]{Dutton10a}, while haloes with deeper
potential wells (and higher escape velocities) will be less
susceptible to feedback mechanisms that suppress star formation. The
exact reason why this results in a preferable scale of $0.4 R_{200}$
remains to be determined.

%% FIGURE 11
\begin{figure}
\centerline{
\includegraphics[width=0.45\textwidth]{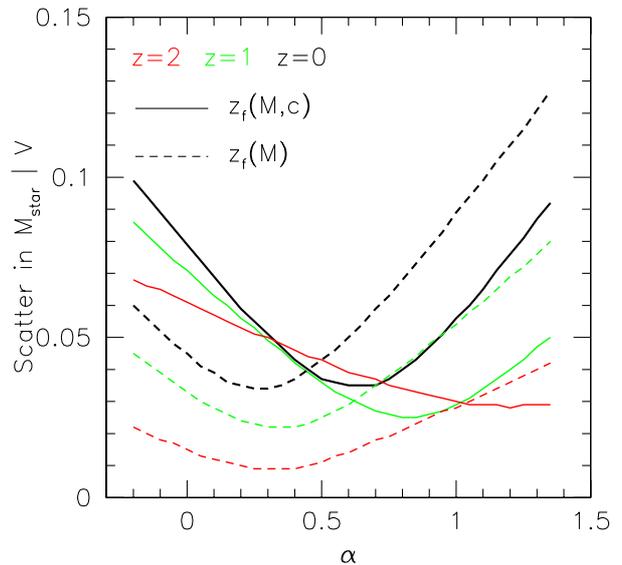}   
}
\caption{Scatter of the stellar mass versus halo velocity relation for
    $V_{\alpha}^{\rm DMO}=V_{200}(V_{\rm max}/V_{200})^{\alpha}$ for
    two halo models at three redshifts ($z=0,1,2$).  The standard
    model (solid lines) has the halo formation redshift $z_{\rm f}$ a
    function of $z=0$ halo mass and concentration, while the control
    halo model has the halo formation redshift solely a function of
    $z=0$ halo mass.}
    \label{fig:vm_scatter4}
\end{figure}

As a first attempt to disentangle these two effects, we run the SAM
again but without a coupling between the halo concentration and mass
accretion history. Here we include only scatter in halo concentration
to make the minimum more apparent. The results are shown in
Fig.~\ref{fig:vm_scatter4}. In the control SAM (dashed lines) the
formation redshift of the halo is solely a function of the halo mass,
$z_{\rm f}(M)$, whereas in the standard SAM (solid lines) the
formation redshift is a function of the mass and concentration at
$z=0$, $z_{\rm f}(M,c)$. At $z=0$ (black lines) the control SAM has a
minimum scatter at $\alphamin\sim 0.3$ compared to $\alphamin\sim 0.6$
for the standard SAM. Thus the correlation between concentration and
mass accretion history is required both to match the NIHAO hydro
simulations and the observed $\alphamin\sim 0.6$.

We also show results at redshifts $z=1$ (green) and $z=2$ (red).  At
higher redshifts the new SAM still has a minimum scatter at
$\alphamin\sim 0.3$, but the standard SAM has $\alphamin$ shifting to
higher values.  Thus we do not expect the same halo velocity definition
to minimize the scatter in stellar mass at all redshifts.

%% FIGURE 12
\begin{figure}
\centerline{
\includegraphics[width=0.45\textwidth]{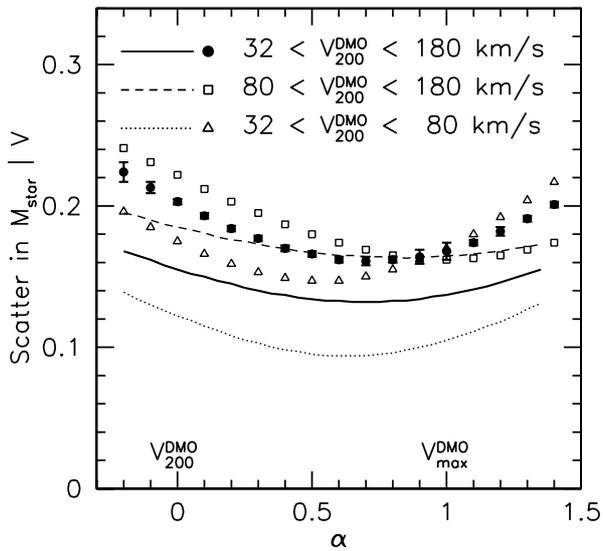}     
}
\caption{Scatter of the stellar mass versus halo velocity relation for
    $V_{\alpha}^{\rm DMO}=V_{200}(V_{\rm max}/V_{200})^{\alpha}$ for
    different halo velocity ranges as indicated. The minimum scatter
    occurs at higher $\alpha$ for higher velocity haloes for both the
    NIHAO hydro simulations (points) and the SAM (lines). }
    \label{fig:vm_scatter5}
\end{figure}

\subsection{Dependence of $\alphamin$ on halo velocity}
Fig.~\ref{fig:vm_scatter5} shows the scatter in the stellar mass versus
$V_{\alpha}$ relation for different halo velocity ranges. Low velocity
haloes ($32 < V_{200}^{\rm DMO} < 80 \kms$) are shown with triangles
(NIHAO) and a dotted line (SAM), and have $\alphamin \sim 0.5$.  High
velocity haloes ($80 < V_{200}^{\rm DMO} < 180 \kms$) are shown with
squares (NIHAO) and a dashed line (SAM), and have $\alphamin \sim
0.9$.

For the SAM we can break the velocity range down further since we have
larger samples.  We find a critical scale of $V_{200}^{\rm DMO} \lta
120 \kms$ ($6\times 10^{11}\Msun$). Below this scale $\alphamin\sim
0.5$ for all haloes we study (down to $V_{200}^{\rm DMO}=25
\kms$). Above this scale $\alphamin$ increases with halo mass reaching
$\alphamin=1$ at $V_{200}^{\rm DMO}\sim 160 \kms$, and $\alphamin=1.4$
at $V_{200}^{\rm DMO}\sim 200 \kms$.  In the control SAM we find the
same critical halo velocity but with $\alphamin\sim 0.3$ for
$V_{200}^{\rm DMO} < 120 \kms$,  while for higher velocities
$\alphamin$ {\it decreases} and approaches $\alphamin=0$ for
$V_{200}^{\rm DMO}=200 \kms$.

In the SAM we identify this critical scale of $V_{200}^{\rm DMO}=120
\kms$ with the threshold for hot halo formation. Below this scale
cooling is very efficient, so that essentially all gas that enters the
halo reaches the central galaxy in a free fall time. Above this scale
cooling starts to become inefficient which reduces the supply of gas
into the central galaxy and lowers subsequent star formation.

\subsection{Dependence of $\alphamin$ on galaxy mass definition}
\begin{figure}
  \centerline{
    \includegraphics[width=0.45\textwidth]{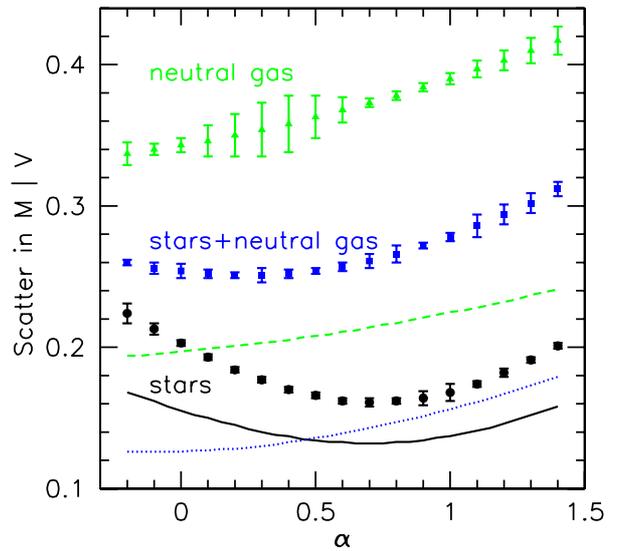}
    }
  \caption{Scatter of the mass versus halo velocity relation for
    $V_{\alpha}^{\rm DMO}=V_{200}(V_{\rm max}/V_{200})^{\alpha}$, and
    three different galaxy mass definitions: neutral gas (green),
    stars (black), and stars + neutral gas (blue). NIHAO simulations
    are shown with points, SAM is shown with lines.}
    \label{fig:vm_scatter6}
\end{figure}

Fig.~\ref{fig:vm_scatter6} shows the scatter in mass at fixed velocity
for three different mass components: stars (black), neutral gas
(green), and stars plus neutral gas (blue). In contrast to the stars,
the neutral gas has $\alphamin\sim 0$, which corresponds to
$V_{200}$. Recall we saw a similar result in Fig.~\ref{fig:mvall}
which showed that neutral gas is better correlated with $V_{200}^{\rm
  DMO}$ than $V_{\rm max}^{\rm DMO}$. As expected, the sum of stars
and neutral gas has intermediate scatter between the two components,
and \alphamin  occurs at an intermediate value of $\alpha$. Again the
hydro sims and SAM give similar results for $\alphamin$ for different
mass components.

%% FIGURE 13

%%%%%%%%%%%%%%%%%%%%%%%%%%%%%%%%%%%%%%%%%%%%%%%%%%%%%%%%%%%%%%%%%%%%%%
%% SECTION 4: SUMMARY
%%%%%%%%%%%%%%%%%%%%%%%%%%%%%%%%%%%%%%%%%%%%%%%%%%%%%%%%%%%%%%%%%%%%%%
\section{Summary}
\label{sec:sum}

We study the scaling relations between galaxy mass and circular
velocity in haloes of mass $10^{10}\lta M_{200} \lta 10^{12}\Msun$. We
use a sample of 83 fully cosmological galaxy formation simulations
from the NIHAO project \citep{Wang15}, and the SAM of
\citet{Dutton12}. We summarize our results as follows:

\begin{itemize}
\item The simulations are consistent with the observed stellar,
  neutral gas, and BTF relations (Fig.~\ref{fig:mass_vflat}).

\item  For the BTF relation, our simulations have a
  small scatter of 0.08-0.14 dex in mass for $70\lta \VHI \lta 240
  \kms$, consistent with observational estimates
  (Fig.~\ref{fig:btf_scatter}). At lower velocities $20\lta \VHI \lta
  70 \kms$ our simulations predict larger scatters of 0.2 to 0.25 dex,
  potentially in conflict with observations.

\item The scatter in stellar mass at fixed halo velocity is constant
  for $30 < V < 180 \,\kms$. The maximum circular velocity of the DMO
  simulation, $\Vmaxdmo$ provides a better predictor of the stellar mass,
  than the virial velocity of the DMO (or hydro) simulation
  (Fig.~\ref{fig:mvall}). However, for gas and baryonic mass, virial
  velocity is a better predictor than maximum velocity. Thus there is
  no single velocity definition that minimizes the scatter in all
  baryonic components.
  
\item The normalization of the stellar mass versus virial velocity
  relation is correlated with halo concentration.
  These correlations are substantially reduced in the
  stellar mass versus $\Vmaxdmo$ relation (Fig.~\ref{fig:vm_c}).
  
\item Measuring the circular velocity at $\simeq 0.4 R_{200}^{\rm DMO}$
  minimizes the scatter in stellar mass versus halo velocity relation at
  0.15 dex (Fig.~\ref{fig:vm_scatter2}).

\item Defining the halo circular velocity as $V_\alpha=V_{200}^{\rm
  DMO} (V_{\rm max}^{\rm DMO}/V_{200}^{\rm DMO})^{\alpha}$, where
  $\alpha=0$ corresponds to virial velocity and $\alpha=1$ corresponds
  to maximum circular velocity,  we find the scatter is minimized at
  0.16 dex for $\alphamin=0.7$ (Fig.~\ref{fig:vm_scatter3}),
  consistent with clustering based
  constraints \citep{Lehmann17}.

\item In the SAM when we decouple the formation time of the halo with
  its concentration we find $\alphamin\sim 0.3$ for all redshifts
  (Fig.~\ref{fig:vm_scatter4}), while in the standard SAM $\alphamin$
  increases with redshift, reaching $\alphamin=1$ at $z=2$. Thus the
  correlation between halo formation time and concentration is
  essential in order to reproduce the $\alphamin\sim0.6$ found in the
  NIHAO hydro sims and from clustering constraints.

\item We find that $\alphamin$ is higher in higher velocity haloes
  (Fig.~\ref{fig:vm_scatter5}). Using the SAM we find this is related
  to the formation of hot haloes above $V_{200}^{\rm DMO} = 120 \kms$
  $(M_{200}^{\rm DMO}=6 \times 10^{11}\Msun)$.

\end{itemize}

The small scatter in the TF relations from the NIHAO
simulations and the SAM of \citet{Dutton12} point to
the simplicity of galaxy formation due to the self-regulation between
star formation and energy feedback from massive stars.

The dependence of scatter in the stellar mass vs halo velocity with
concentration appears to be a fundamental property of galaxy formation
in a \LCDM universe, and thus can be used to improve the accuracy of
halo abundance matching models.

\section*{Acknowledgements} 

We thank the referee for useful comments that improved the
presentation and motivated deeper analysis. This research was carried
out on the High Performance Computing resources at New York University
Abu Dhabi; on the {\sc theo} cluster of the Max-Planck-Institut f\"ur
Astronomie and the {\sc hydra} cluster at the Rechenzentrum in
Garching; and the Milky Way supercomputer, which is funded by the
Deutsche Forschungsgemeinschaft (DFG) through Collaborative Research
Center (SFB 881) ``The Milky Way System'' (subproject Z2) and  hosted
and co-funded by the J\"ulich Supercomputing Center (JSC). We greatly
appreciate the contributions of all these computing allocations.
TB and TAG were supported by the Sonderforschungsbereich SFB 881 ``The Milky
Way System'' (subprojects A1 and A2) of the DFG.
The analysis made use of the {\sc pynbody} package \citep{Pontzen13}.
XK acknowledges the support from NSFC project No.11333008 and the
Strategic Priority Research Program `The Emergence of Cosmological
Structures' of the CAS(No.XD09010000).

%%%%%%%%%%%%%%%%%%%%%%%%%%%%%%%%%%%%%%%%%%%%%%%%%%%%%%%%%%%%%%%%%%%%%%
%%  REFERENCES
%%%%%%%%%%%%%%%%%%%%%%%%%%%%%%%%%%%%%%%%%%%%%%%%%%%%%%%%%%%%%%%%%%%%%% 


%\bibliographystyle{mnras}


\begin{table*}
\begin{center}
  \caption{Data from NIHAO simulations from Fig.~\ref{fig:mvall}. All
    logarithms are base 10. Sizes are in units of $\kpc$, velocities
    in $\kms$, and masses in $\Msun$. Column (1) NIHAO simulation ID,
    column (2) \HI\, radius, column (3) circular velocity measured at
    the \HI\, radius, column (4) maximum circular velocity of the dark
    matter only simulation, column (5) virial circular velocity from
    the hydro simulation, column (6) virial circular velocity from the
    dark matter only simulation, column (7) stellar mass, column (8)
    neutral gas mass, column (9) galaxy baryonic mass (column 7 plus
    column 8), column (10) baryonic mass inside the virial radius of
    the hydro simulation, column (11) baryonic mass associated with
    the dark matter only simulation, $M_{\rm bar}^{\rm
      DMO}\equiv(\Omegab/\Omegam)M_{200}^{\rm DMO}$.}
\label{tab:mvall}  
\begin{tabular}{ccccccccccc}
\hline
ID & $\log(R_{\rm HI})$ & $\log(V_{\rm HI})$ & $\log(V_{\rm max}^{\rm DMO})$ & $\log(V_{200})$ & $\log(V_{200}^{\rm DMO})$ &  $\log(M_{\rm star})$ & $\log(M_{\rm neut})$ & $\log(M_{\rm gal})$ & $\log(M_{\rm bar})$ & $\log(M_{\rm bar}^{\rm DMO})$ \\
 & [$\kpc$] & [$\kms$] & [$\kms$] & [$\kms$] & [$\kms$] & [$\Msun$] & [$\Msun$] & [$\Msun$] & [$\Msun$] & [$\Msun$] \\
(1) & (2) & (3) & (4) & (5) & (6) & (7) & (8) & (9) & (10) & (11)\\
\hline 
g4.36e09 & -0.032 &  1.318 &  1.579 &  1.483 &  1.492 &  4.518 &  6.574 &  6.578 &  8.222 &  9.205 \\ 
g4.99e09 & -0.086 &  1.351 &  1.544 &  1.406 &  1.445 &  5.575 &  7.244 &  7.253 &  7.926 &  9.064 \\ 
g5.22e09 & -0.036 &  1.394 &  1.547 &  1.420 &  1.455 &  5.070 &  7.164 &  7.167 &  7.961 &  9.094 \\ 
g5.41e09 & -0.244 &  1.319 &  1.570 &  1.390 &  1.438 &  6.084 &  7.044 &  7.089 &  7.627 &  9.043 \\ 
g5.59e09 & -0.208 &  1.329 &  1.576 &  1.424 &  1.471 &  6.240 &  7.184 &  7.231 &  7.860 &  9.142 \\ 
g5.84e09 & -0.108 &  1.316 &  1.555 &  1.412 &  1.443 &  5.064 &  7.214 &  7.217 &  7.961 &  9.056 \\ 
g7.05e09 & -0.004 &  1.313 &  1.612 &  1.494 &  1.551 &  6.355 &  6.484 &  6.725 &  8.191 &  9.381 \\ 
g7.34e09 & -0.187 &  1.330 &  1.558 &  1.414 &  1.454 &  5.626 &  7.074 &  7.089 &  7.900 &  9.089 \\ 
g9.26e09 &  0.097 &  1.404 &  1.535 &  1.416 &  1.470 &  4.730 &  6.884 &  6.887 &  8.052 &  9.138 \\ 
g1.09e10 &  0.787 &  1.603 &  1.622 &  1.501 &  1.523 &  6.810 &  8.474 &  8.483 &  8.786 &  9.298 \\ 
g1.18e10 &  0.068 &  1.441 &  1.624 &  1.500 &  1.541 &  6.531 &  7.524 &  7.566 &  8.312 &  9.352 \\ 
g1.23e10 &  0.170 &  1.413 &  1.605 &  1.473 &  1.524 &  6.201 &  6.344 &  6.580 &  8.197 &  9.300 \\ 
g1.44e10 &  0.610 &  1.679 &  1.676 &  1.563 &  1.596 &  6.826 &  8.884 &  8.888 &  9.042 &  9.517 \\ 
g1.50e10 &  0.107 &  1.477 &  1.655 &  1.536 &  1.573 &  6.531 &  7.384 &  7.441 &  8.332 &  9.446 \\ 
g1.57e10 & -0.056 &  1.405 &  1.695 &  1.524 &  1.563 &  6.950 &  7.434 &  7.557 &  8.216 &  9.418 \\ 
g1.88e10 &  0.562 &  1.622 &  1.715 &  1.578 &  1.618 &  7.215 &  8.224 &  8.264 &  8.418 &  9.583 \\ 
g1.89e10 &  0.427 &  1.560 &  1.693 &  1.597 &  1.623 &  7.108 &  8.234 &  8.265 &  9.073 &  9.596 \\ 
g1.90e10 &  0.500 &  1.538 &  1.684 &  1.614 &  1.645 &  7.077 &  8.154 &  8.189 &  8.891 &  9.662 \\ 
g1.92e10 &  0.297 &  1.503 &  1.656 &  1.586 &  1.618 &  6.721 &  7.904 &  7.931 &  8.987 &  9.582 \\ 
g1.95e10 &  0.049 &  1.443 &  1.665 &  1.533 &  1.565 &  6.581 &  7.634 &  7.671 &  8.477 &  9.422 \\ 
g2.09e10 &  0.986 &  1.630 &  1.683 &  1.560 &  1.590 &  6.914 &  7.924 &  7.964 &  8.772 &  9.500 \\ 
g2.34e10 &  0.367 &  1.483 &  1.737 &  1.623 &  1.654 &  7.133 &  7.344 &  7.552 &  8.767 &  9.690 \\ 
g2.39e10 &  0.193 &  1.462 &  1.661 &  1.541 &  1.575 &  6.772 &  7.714 &  7.761 &  8.618 &  9.454 \\ 
g2.63e10 &  0.158 &  1.551 &  1.803 &  1.630 &  1.671 &  7.631 &  7.344 &  7.812 &  8.555 &  9.741 \\ 
g2.64e10 &  0.990 &  1.759 &  1.749 &  1.658 &  1.682 &  7.459 &  9.264 &  9.271 &  9.402 &  9.775 \\ 
g2.80e10 &  0.549 &  1.634 &  1.777 &  1.634 &  1.703 &  7.565 &  8.504 &  8.551 &  9.058 &  9.837 \\ 
g2.83e10 &  0.835 &  1.733 &  1.759 &  1.621 &  1.657 &  7.481 &  8.994 &  9.007 &  9.161 &  9.699 \\ 
g2.94e10 &  0.348 &  1.574 &  1.793 &  1.656 &  1.689 &  7.764 &  7.684 &  8.027 &  8.812 &  9.797 \\ 
g3.19e10 &  0.352 &  1.554 &  1.793 &  1.665 &  1.699 &  7.167 &  7.994 &  8.054 &  9.212 &  9.826 \\ 
g3.44e10 &  1.128 &  1.754 &  1.799 &  1.719 &  1.759 &  7.797 &  8.874 &  8.909 &  9.644 & 10.005 \\ 
g3.67e10 &  0.484 &  1.657 &  1.805 &  1.653 &  1.692 &  7.739 &  7.704 &  8.023 &  8.759 &  9.803 \\ 
g3.93e10 &  0.852 &  1.674 &  1.733 &  1.659 &  1.675 &  7.567 &  8.974 &  8.991 &  9.355 &  9.755 \\ 
g4.27e10 &  0.925 &  1.724 &  1.797 &  1.697 &  1.711 &  7.767 &  8.884 &  8.916 &  9.471 &  9.863 \\ 
g4.48e10 &  0.849 &  1.755 &  1.837 &  1.748 &  1.771 &  8.122 &  8.924 &  8.988 &  9.476 & 10.042 \\ 
g4.86e10 &  0.922 &  1.873 &  1.886 &  1.725 &  1.751 &  8.082 &  9.364 &  9.386 &  9.471 &  9.981 \\ 
g4.94e10 &  0.648 &  1.743 &  1.867 &  1.728 &  1.759 &  8.033 &  8.524 &  8.645 &  9.268 & 10.005 \\ 
g4.99e10 &  0.818 &  1.752 &  1.840 &  1.717 &  1.761 &  8.078 &  8.834 &  8.904 &  9.296 & 10.011 \\ 
g5.05e10 &  0.580 &  1.724 &  1.874 &  1.698 &  1.771 &  7.971 &  8.274 &  8.449 &  9.051 & 10.041 \\ 
g6.12e10 &  1.036 &  1.816 &  1.869 &  1.719 &  1.760 &  7.954 &  9.014 &  9.050 &  9.424 & 10.007 \\ 
g6.37e10 &  1.236 &  1.875 &  1.877 &  1.839 &  1.852 &  8.310 &  9.674 &  9.692 & 10.043 & 10.283 \\ 
g6.77e10 &  0.958 &  1.824 &  1.885 &  1.810 &  1.830 &  8.679 &  9.394 &  9.470 &  9.872 & 10.219 \\ 
g6.91e10 &  0.876 &  1.858 &  1.931 &  1.770 &  1.803 &  8.396 &  8.734 &  8.898 &  9.322 & 10.138 \\ 
g6.96e10 &  1.120 &  1.860 &  1.915 &  1.805 &  1.861 &  8.552 &  9.224 &  9.308 &  9.741 & 10.313 \\ 
g8.89e10 &  0.931 &  1.897 &  1.915 &  1.809 &  1.845 &  8.601 &  9.414 &  9.476 &  9.794 & 10.262 \\ 
g9.59e10 &  1.122 &  1.853 &  1.892 &  1.803 &  1.845 &  8.419 &  9.594 &  9.622 &  9.944 & 10.264 \\ 
g1.05e11 &  1.083 &  1.908 &  1.939 &  1.844 &  1.870 &  8.744 &  9.554 &  9.616 &  9.932 & 10.338 \\ 
g1.08e11 &  1.108 &  1.966 &  1.984 &  1.847 &  1.886 &  8.922 &  9.544 &  9.637 &  9.845 & 10.387 \\ 
g1.37e11 &  0.876 &  1.990 &  2.036 &  1.877 &  1.908 &  9.303 &  9.284 &  9.594 &  9.787 & 10.452 \\ 
g1.52e11 &  1.205 &  1.982 &  2.007 &  1.886 &  1.906 &  8.882 &  9.764 &  9.817 & 10.179 & 10.448 \\ 
g1.57e11 &  1.040 &  1.926 &  1.986 &  1.886 &  1.914 &  9.053 &  9.714 &  9.800 & 10.050 & 10.471 \\ 
g1.59e11 &  1.338 &  1.977 &  1.979 &  1.896 &  1.911 &  8.813 &  9.924 &  9.956 & 10.263 & 10.461 \\ 
g1.64e11 &  1.184 &  1.990 &  1.965 &  1.917 &  1.936 &  8.944 & 10.044 & 10.077 & 10.385 & 10.536 \\ 

\hline
\end{tabular}
\end{center}
\end{table*}

\setcounter{table}{0}
\begin{table*}
\begin{center}
  \caption{-- continued.}
\label{tab:fits}  
\begin{tabular}{ccccccccccc}
\hline
ID & $\log(R_{\rm HI})$ & $\log(V_{\rm HI})$ & $\log(V_{\rm max}^{\rm DMO})$ & $\log(V_{200})$ & $\log(V_{200}^{\rm DMO})$ &  $\log(M_{\rm star})$ & $\log(M_{\rm neut})$ & $\log(M_{\rm gal})$ & $\log(M_{\rm bar})$ & $\log(M_{\rm bar}^{\rm DMO})$ \\
(1) & (2) & (3) & (4) & (5) & (6) & (7) & (8) & (9) & (10) & (11)\\
\hline 
g2.04e11 &  0.915 &  2.065 &  2.060 &  1.927 &  1.955 &  9.670 &  9.684 &  9.978 & 10.139 & 10.593 \\ 
g2.19e11 &  0.980 &  1.917 &  1.949 &  1.860 &  1.886 &  8.961 &  9.564 &  9.661 & 10.039 & 10.387 \\ 
g2.39e11 &  0.876 &  2.103 &  2.062 &  1.958 &  1.982 &  9.762 &  9.744 & 10.054 & 10.248 & 10.675 \\ 
g2.41e11 &  0.981 &  2.060 &  2.095 &  1.955 &  1.998 &  9.610 &  9.804 & 10.019 & 10.232 & 10.723 \\ 
g2.42e11 &  0.907 &  2.090 &  2.110 &  1.963 &  1.997 &  9.736 &  9.614 &  9.980 & 10.120 & 10.720 \\ 
g2.54e11 &  0.931 &  2.057 &  2.063 &  1.963 &  2.009 &  9.540 &  9.884 & 10.046 & 10.305 & 10.756 \\ 
g2.57e11 &  1.157 &  2.147 &  2.095 &  1.976 &  2.000 & 10.033 &  9.904 & 10.274 & 10.433 & 10.729 \\ 
g3.06e11 &  0.930 &  2.124 &  2.095 &  1.985 &  2.017 &  9.868 &  9.874 & 10.172 & 10.392 & 10.779 \\ 
g3.21e11 &  1.080 &  2.056 &  2.039 &  1.981 &  1.998 &  9.563 &  9.944 & 10.095 & 10.453 & 10.721 \\ 
g3.23e11 &  1.141 &  1.869 &  1.902 &  1.804 &  1.848 &  8.544 &  9.644 &  9.677 &  9.953 & 10.272 \\ 
g3.49e11 &  1.206 &  2.097 &  2.121 &  2.030 &  2.048 &  9.592 & 10.164 & 10.267 & 10.607 & 10.872 \\ 
g3.55e11 &  1.318 &  2.112 &  2.091 &  2.029 &  2.039 &  9.575 & 10.324 & 10.395 & 10.635 & 10.846 \\ 
g3.59e11 &  1.143 &  2.077 &  2.066 &  2.002 &  2.017 &  9.635 & 10.164 & 10.276 & 10.539 & 10.778 \\ 
g3.61e11 &  1.172 &  2.226 &  2.172 &  2.032 &  2.061 & 10.333 &  9.964 & 10.487 & 10.567 & 10.910 \\ 
g3.71e11 &  1.426 &  2.145 &  2.125 &  2.024 &  2.042 & 10.073 & 10.064 & 10.369 & 10.626 & 10.855 \\ 
g5.02e11 &  1.356 &  2.188 &  2.153 &  2.074 &  2.093 & 10.162 & 10.294 & 10.534 & 10.731 & 11.008 \\ 
g5.31e11 &  1.387 &  2.183 &  2.156 &  2.062 &  2.087 & 10.214 &  9.914 & 10.391 & 10.716 & 10.989 \\ 
g5.36e11 &  1.116 &  2.193 &  2.182 &  2.101 &  2.138 & 10.063 & 10.284 & 10.488 & 10.816 & 11.143 \\ 
g5.38e11 &  1.203 &  2.217 &  2.183 &  2.091 &  2.106 & 10.265 & 10.154 & 10.514 & 10.749 & 11.046 \\ 
g5.46e11 &  1.161 &  2.080 &  2.075 &  1.991 &  2.040 &  9.572 & 10.064 & 10.185 & 10.446 & 10.848 \\ 
g5.55e11 &  1.003 &  2.214 &  2.135 &  2.059 &  2.091 & 10.232 & 10.024 & 10.441 & 10.650 & 11.002 \\ 
g6.96e11 &  1.137 &  2.278 &  2.206 &  2.122 &  2.159 & 10.513 & 10.374 & 10.750 & 10.936 & 11.207 \\ 
g7.08e11 &  1.259 &  2.291 &  2.208 &  2.123 &  2.138 & 10.482 & 10.514 & 10.799 & 10.972 & 11.143 \\ 
g7.44e11 &  1.212 &  2.257 &  2.210 &  2.178 &  2.168 & 10.253 & 10.754 & 10.873 & 11.157 & 11.232 \\ 
g7.55e11 &  1.342 &  2.276 &  2.209 &  2.138 &  2.160 & 10.490 & 10.604 & 10.852 & 11.017 & 11.208 \\ 
g7.66e11 &  1.036 &  2.365 &  2.228 &  2.143 &  2.155 & 10.771 & 10.074 & 10.850 & 11.006 & 11.194 \\ 
g8.06e11 &  1.124 &  2.306 &  2.206 &  2.146 &  2.152 & 10.647 & 10.244 & 10.791 & 11.043 & 11.185 \\ 
g8.13e11 &  1.568 &  2.308 &  2.265 &  2.152 &  2.168 & 10.824 & 10.304 & 10.939 & 11.060 & 11.231 \\ 
g8.26e11 &  1.437 &  2.322 &  2.265 &  2.157 &  2.169 & 10.669 & 10.644 & 10.957 & 11.088 & 11.236 \\ 
g8.28e11 &  1.288 &  2.270 &  2.218 &  2.176 &  2.172 & 10.245 & 10.684 & 10.819 & 11.152 & 11.244 \\ 
g1.12e12 &  1.237 &  2.367 &  2.260 &  2.170 &  2.185 & 10.894 &  9.904 & 10.937 & 11.138 & 11.282 \\ 

\hline
\end{tabular}
\end{center}
\end{table*}

% Don't change these lines
\bsp	% typesetting comment
\label{lastpage}
\end{document}